%% file: manuscript.tex
\documentclass[acmsmall,nonacm]{acmart}

\AtBeginDocument{%
  }

\setcopyright{acmlicensed}
\copyrightyear{2018}
\acmYear{2018}
\acmDOI{XXXXXXX.XXXXXXX}

\acmConference[Conference acronym 'XX]{Make sure to enter the correct
  conference title from your rights confirmation emai}{June 03--05,
  2018}{Woodstock, NY}
\acmISBN{978-1-4503-XXXX-X/18/06}

\PassOptionsToPackage{table,xcdraw}{xcolor}
\usepackage[table,xcdraw]{xcolor} 
\usepackage{booktabs}   
\usepackage{textcomp}
\usepackage{graphicx}  
\usepackage{subfigure} 
\usepackage{subcaption}

\usepackage{xcolor}
\usepackage{array}
\usepackage{colortbl}

\usepackage{geometry}
\usepackage{pdflscape}

\usepackage[normalem]{ulem} 

\definecolor{lightgray}{gray}{0.9}
\definecolor{darkblue}{rgb}{0.0, 0.0, 0.5}
\definecolor{headercolor}{rgb}{0.3, 0.5, 0.7} 

\begin{document}

\title{Personalizing Emotion-aware Conversational Agents? Exploring User Traits-driven Conversational Strategies for Enhanced Interaction}


\author{Yuchong Zhang}
\orcid{0000-0003-1804-6296}
\authornote{Contributed equally to this work.}
\authornote{Corresponding author.}
\email{yuchongz@kth.se}
\affiliation{%
  \institution{KTH Royal Institute of Technology}
  \city{Stockholm}
  \country{Sweden}
}

\author{Yong Ma}
\orcid{0000-0002-8398-4118}
\authornotemark[1] 
\affiliation{%
  \institution{University of Bergen}
  \city{Bergen}
  \country{Norway}}
\email{yong.ma@uib.no}

\author{Di Fu}
\orcid{0000-0002-5385-2982}
\email{d.fu@surrey.ac.uk}
\affiliation{%
  \institution{University of Surrey}
  \city{Guildford}
  \country{United Kingdom}
}

\author{Stephanie Zubicueta Portales}
\orcid{0009-0004-2459-1949}
\affiliation{%
  \institution{Norwegian University of Science and Technology}
  \city{Trondheim}
  \country{Norway}}
\email{stephanie.portales@student.uib.no}

\author{Morten Fjeld}
\orcid{0000-0002-9562-5147}
\affiliation{%
  \institution{University of Bergen}
  \city{Bergen}
  \country{Norway}}
\email{morten.fjeld@uib.no}
\additionalaffiliation{%
  \institution{Chalmers University of Technology}
  \city{Gothenburg}
  \country{Sweden}}
\email{fjeld@chalmers.se}

\author{Danica Kragic}
\orcid{0000-0003-2965-2953}
\affiliation{%
  \institution{KTH Royal Institute of Technology}
  \city{Stockholm}
  \country{Sweden}}
\email{dani@kth.se}

\renewcommand{\shortauthors}{Zhang and Ma et al.}

\begin{abstract}
Conversational agents (CAs) are increasingly embedded in daily life, yet their ability to navigate user emotions \textcolor{black}{efficiently} is still evolving. This study investigates how users with varying traits -- gender, personality, and cultural background -- adapt their interaction strategies with emotion-aware CAs in specific emotional scenarios. Using an emotion-aware CA prototype \textcolor{black}{expressing five distinct emotions (\textit{neutral}, \textit{happy}, \textit{sad}, \textit{angry}, and \textit{fear}) through male and female voices}, we examine how interaction dynamics shift across different voices and emotional contexts \textcolor{black}{through empirical studies}. Our findings reveal distinct variations in user engagement \textcolor{black}{and conversational strategies} based on individual traits, emphasizing the value of personalized, emotion-sensitive interactions. \textcolor{black}{By analyzing both qualitative and quantitative data,} we demonstrate that tailoring CAs to user characteristics can enhance user satisfaction and interaction quality. This work underscores the critical need for ongoing research to design CAs that not only recognize but also adaptively respond to emotional needs, ultimately supporting a diverse user groups more effectively.

\end{abstract}




\received{20 February 2007}
\received[revised]{12 March 2009}
\received[accepted]{5 June 2009}

\maketitle


\input{Sections/Introduction}

\input{Sections/Related_Work}
\input{Sections/Research_Design}

\input{Sections/Experiment}

\input{Sections/Quantitative_results}

\input{Sections/Qualitative_Results}

\input{Sections/Key_Findings}
\input{Sections/Discussion}
\input{Sections/Conclusion}

\begin{acks}
To Robert, for the bagels and explaining CMYK and color spaces.
\end{acks}

\bibliographystyle{ACM-Reference-Format}
\bibliography{sample-base}

\appendix
\label{appen}

\clearpage

\section{The Smiley Face Icons for the Five Emotions}
\label{appen.a}

\textcolor{black}{We only present the computer views of the male-voiced CAs, as the interfaces for the female-voiced CAs are identical.}

\begin{figure*}[!h]
    \centering
    \includegraphics[width=\linewidth]{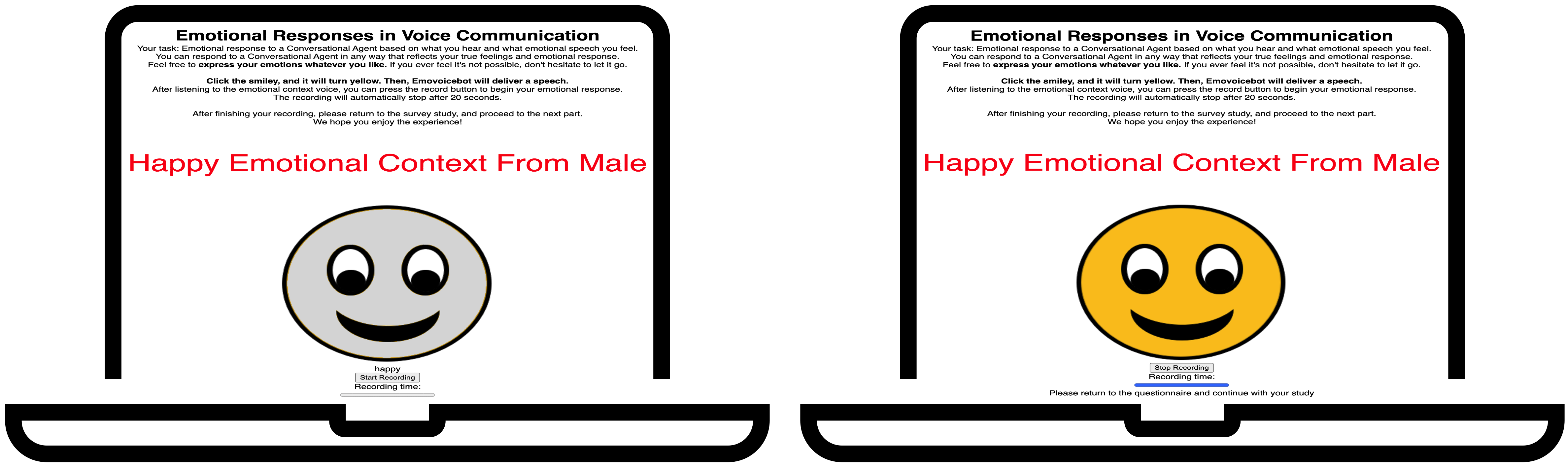}
    \caption{Computer views of interacting with the CAs with \textit{\textbf{happy}} emotion with the male voice. Left: before the conversation; right: after the conversation.}
\end{figure*}

\begin{figure*}[!h]
    \centering
    \includegraphics[width=\linewidth]{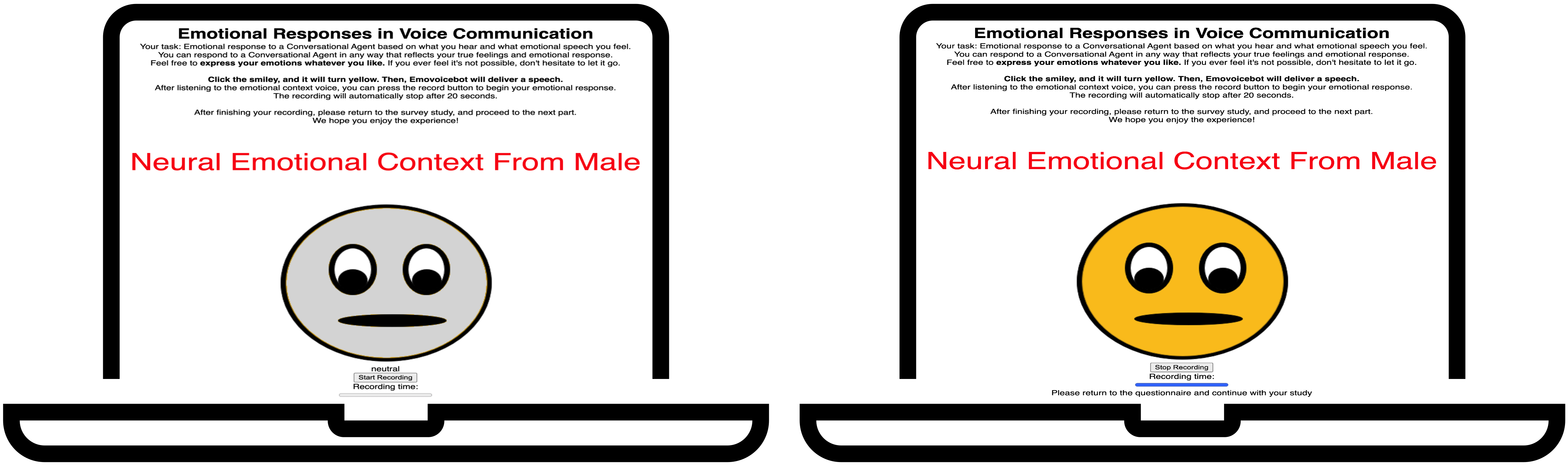}
    \caption{Computer views of interacting with the CAs with \textit{\textbf{neutral}} emotion with the male voice. Left: before the conversation; right: after the conversation.}
\end{figure*}

\begin{figure*}[!h]
    \centering
    \includegraphics[width=\linewidth]{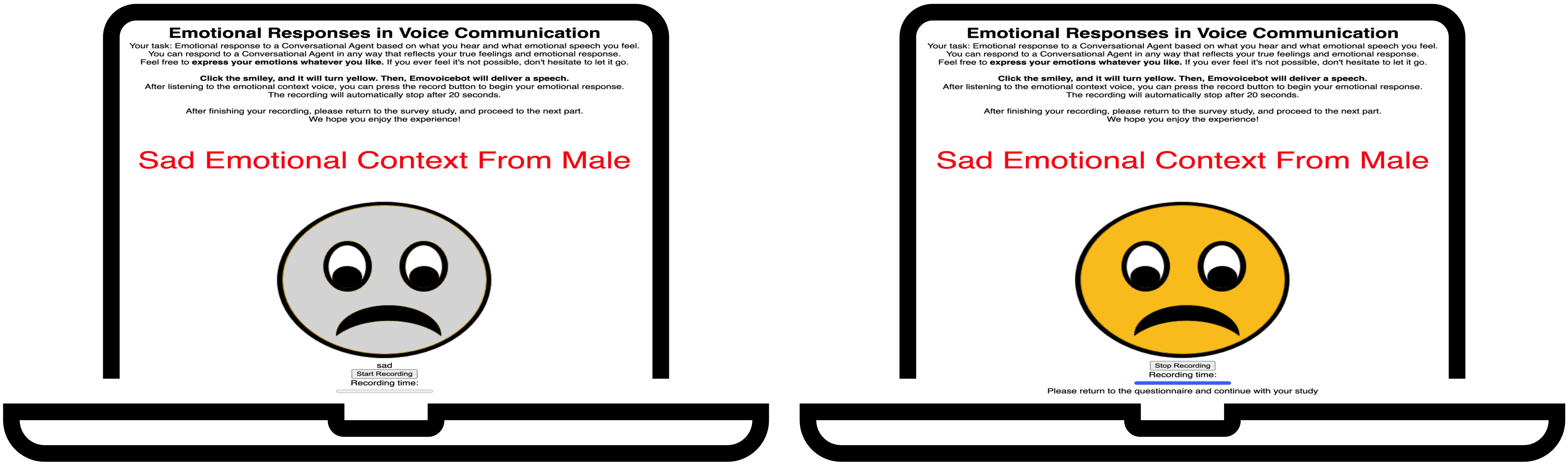}
    \caption{Computer views of interacting with the CAs with \textit{\textbf{sad}} emotion with the male voice. Left: before the conversation; right: after the conversation.}
\end{figure*}

\begin{figure*}[!h]
    \centering
    \includegraphics[width=\linewidth]{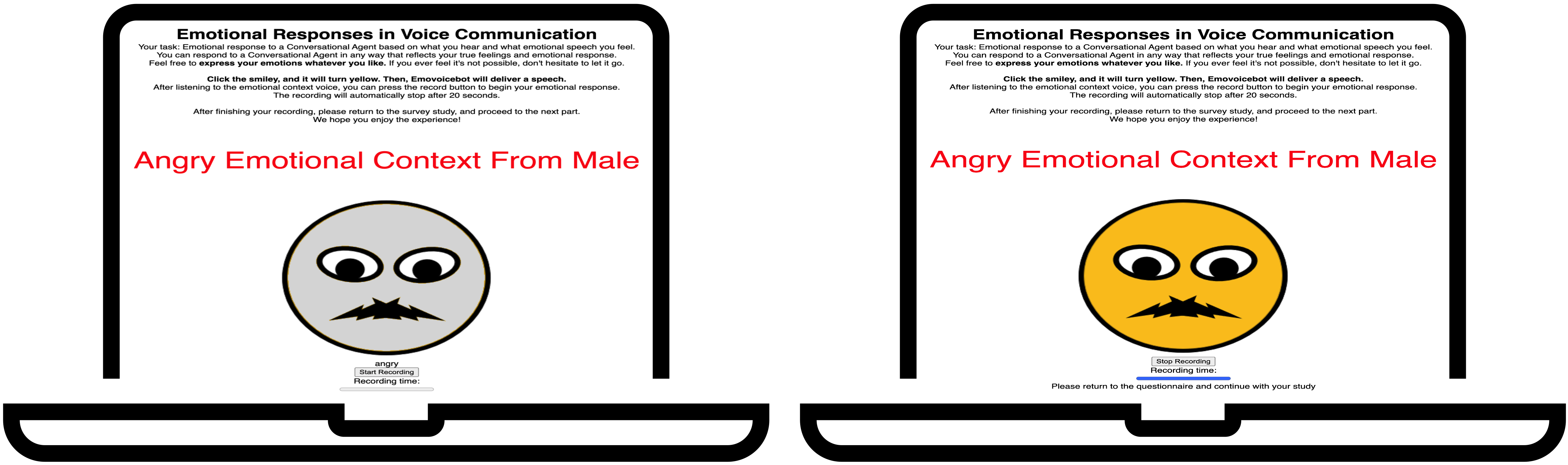}
    \caption{Computer views of interacting with the CAs with \textit{\textbf{angry}} emotion with the male voice. Left: before the conversation; right: after the conversation.}
\end{figure*}

\begin{figure*}[!h]
    \centering
    \includegraphics[width=\linewidth]{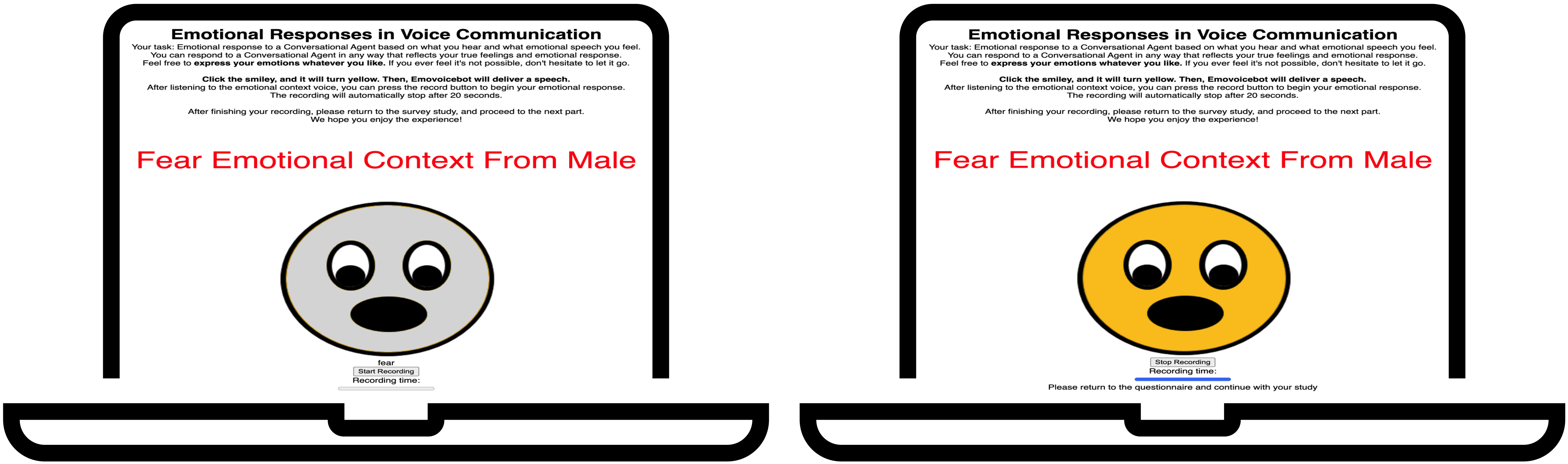}
    \caption{Computer views of interacting with the CAs with \textit{\textbf{fear}} emotion with the male voice. Left: before the conversation; right: after the conversation.}
\end{figure*}

\clearpage

\section{Scripts of Emotional Scenarios}
\label{appen.b}

\textcolor{black}{To simulate realistic interactions and evaluate user engagement with CAs, we designed five emotional scenarios: neutral, happy, sad, angry, and fear. Each scenario was carefully scripted in our CAs to align with the intended emotional state, incorporating contextually relevant language and tone to ensure authenticity and engagement. The development of these scripts was the result of a collaborative discussion among all authors based on previous literature, while one of our authors, who is also a psychologist, offered expert insights and recommendations. Following these discussions, we finalised five scripts of the emotional scenarios which are described in detail below:}

\textcolor{black}{
\begin{itemize}
        \item \textbf{Neutral:} "Hi, I am Emovoice bot. I’ll share a story with you later. This morning, I put on my shoes before leaving the house. It was just part of my usual routine."
        \item \textbf{Happy:} "Hi there! I'm Emovoice bot, and guess what? I've got something super exciting to share with you! Tomorrow, I’m going on an adventure to my absolute favourite city! I can’t wait to explore, see the sights, and maybe even try some new things. It’s going to be amazing! I’ll be sure to tell you all about it when I get back. Are you as excited as I am?"
	\item \textbf{Sad:} "Hi, I am Emovoice bot. I wanted to share something that's been weighing on my mind. I saw the news about children suffering in the war,     and it made me feel really sad. It’s hard to see so much pain and not be affected by it. Could you comfort me? I just need to talk about it with someone who 
        understands."
        \item \textbf{Angry:}"Hi, I am Emovoice bot. I have something very upsetting to share. I was betrayed by someone I considered a close friend—or even like family! I’m so angry right now. I trusted them, and they let me down in the worst way possible. I don’t know what to do with all this frustration. Can you help me figure out how to deal with this anger?"
        \item \textbf{Fear:}"Hi, I am Emovoice bot. I have a very frightening story to share. I was walking alone around some glaciers when suddenly, I slipped and fell down a crack in the ice. The more I try to move, the deeper I sink into the darkness. I’m so scared right now, I feel so trapped, I’m shaking with fear. Can you help me calm down?"    
\end{itemize}
}



\end{document}

%% file: Sections/Introduction.tex
\section{Introduction}
\label{intro}

Conversational agents (CAs) have become increasingly popular in applications such as survey studies \cite{kim2019comparing,xiao2020tell,ma2024understanding}, education \cite{kumar2010socially,winkler2020sara}, mental healthcare \cite{fitzpatrick2017delivering,lee2019caring,morris2018towards}, and tourism \cite{niculescu2014design,van2021chatbots}. This rising prevalence has spurred research into enhancing the emotional intelligence of CAs, a concept rooted in psychology as the ability to express, manipulate, evaluate, and employ emotions \cite{yang2017perceived,ma2019exploring}. The integration of emotional intelligence into CAs aims to enrich user interactions by making these agents more responsive to emotional cues. As a representative, voice assistants, a common type of CA, are now ubiquitous and available across a range of devices such as smartphones, computers, and smart speakers \cite{clark2019state}, utilized in diverse contexts, including smart home and automotive settings \cite{braun2019your,andrade2023voice,lei2018insecurity}. Moreover, CAs are perceived as particularly appealing if adapted to user preferences, behaviour, and background \cite{cowan2015voice,dahlback2007similarity,dewaele2000personality,foster2010user}. Some researchers confirm that CAs serve beyond simple task execution \cite{zhang2020automated}, functioning as more interactive and adaptable conversational partners to users \cite{clark2019makes}. Their appeal significantly increases when they are tailored to users’ preferences, behaviours, and backgrounds \cite{zargham2022want,yang2021designing}.

\textcolor{black}{Emotional intelligence, defined as the ability to recognize, interpret, and respond appropriately to emotions \cite{salovey1990emotional,ma2023emotion}, is crucial for fostering meaningful human-agent interactions.}
Emotion-aware CAs are designed to detect, interpret, and respond to user emotional states, fostering dynamic and engaging interactions \cite{ma2023emotion,ma2024understanding}. As CAs become more emotionally responsive, they facilitate a two-way exchange where users not only receive emotional cues from the agents \cite{feine2019taxonomy} but also respond with their own emotions, shaping the conversation's direction \cite{candello2017evaluating}. This bidirectional emotional exchange highlights the need to understand how users navigate emotional scenarios, adapting their communication style and behaviour based on the CA’s perceived emotions \cite{aneja2021understanding,hu2022acoustically}. In emotionally charged scenarios \cite{scheirer2002frustrating}, users may employ various strategies, such as providing reassurance, showing empathy, or using humour \cite{mcquiggan2007modeling,overall2014attachment}. These strategies are influenced by user traits, past experiences, and the CA's emotional context. For example, in response to a sad or distressed CA, a user may adopt a comforting tone, while an angry CA might prompt users to either defuse the situation with calm responses or match the intensity of the emotion. Understanding these user strategies is crucial for designing CAs that can effectively interact in emotionally appropriate ways \cite{chaves2018single,sciuto2018hey}, ensuring that the agents not only recognize user emotions but also respond in a manner that aligns with user expectations and conversational goals. This exploration of user responses to emotional CAs opens new avenues for developing adaptive dialogue systems that enhance the quality and effectiveness of human-agent interactions.

While today’s CAs like Siri and Google Assistant are not fully emotion-aware, advancements in AI suggest that CAs will soon recognize and respond to human emotions effectively \cite{gornemann2022emotional,diederich2022design,ma2025advancing}. This evolution raises critical questions about how CAs should respond to provide meaningful, supportive assistance. Designing CAs to navigate complex emotional landscapes thoughtfully can lead to more empathetic, personalized, and contextually relevant interactions. However, challenges remain, including interpreting emotional cues accurately, adapting responses, and maintaining user trust and comfort \cite{cai2022impacts,gupta2022trust,10.2312:evs.20201049}. Understanding user expectations and developing suitable response strategies are key to creating emotionally engaging CAs without crossing boundaries or seeming intrusive.

Despite advances, current research focuses primarily on CA interactions and user perceptions \cite{lee2019does,jeong2019exploring,luger2016like}, while a deeper understanding of user-specific factors influencing emotional interactions remains scarce. Prior work has highlighted the benefits of tailoring language complexity to diverse audience backgrounds, addressing the challenges of balancing simplicity with informational richness~\cite{august2024know}. Many studies, however, treat users as a homogeneous group, neglecting individual differences that significantly affect interaction styles and outcomes \cite{nass2005wired,zhang2021affective}. Adaptive and empathetic CA systems have demonstrated potential in improving user experiences by employing emotion lexicons, embeddings, and reinforcement learning-based feedback loops to generate contextually appropriate and emotionally resonant responses \cite{hu2022acoustically,hertz2021adaptive}. However, these approaches often face limitations, including trade-offs between emotional richness and coherence, and a dependence on static, predefined lexicons. The advent of Large Language Models (LLMs) has propelled research in emotional response generation. Notable advancements include models like MECM, which leverage hierarchical latent variables and emotion classifiers for empathetic multi-turn conversations, and findings that ChatGPT-generated questionnaires can increase user happiness while mitigating sadness \cite{rasool2024emotion, zou2024pilot}. Nonetheless, these approaches remain constrained by small sample sizes, static emotion models, and mismatched user expectations, often falling short of achieving real-time adaptability and personalized emotional depth.

To gain deeper insights into human response strategies when interacting with emotion-aware CAs, we aim to explore how individuals respond to different emotional scenarios presented by the CAs. Moreover, our study takes this exploration a step further by examining users’ preferences and strategies in emotion-aware CA dialogues during different situations. By reflecting on past interactions and envisioning future conversational approaches, we investigate how users adapt their strategies in emotional contexts and explore the relationships between user traits (gender, personality, cultural background) and their chosen interaction strategies with CAs. Through this approach, we seek to uncover how users would prefer CAs to behave in emotionally charged contexts, with the goal of informing future CA design. The research questions were intended to address the following:

\begin{itemize}
    \item[-] \textbf{RQ1:} How do users respond to various emotional contexts when interacting with emotion-aware CAs?
    \item[-] \textbf{RQ2:} What strategies do users with different traits employ when engaging with CAs in specific emotional scenarios?
\end{itemize}

Initial findings suggest that user interaction dynamics, such as conversational strategies with emotion-aware CAs, can differ across user traits such as gender, personality, and cultural background. In addition, distinction in voice (male/female) in CAs and differently expressive emotions can also affect interaction dynamics. While the study’s results are preliminary and limited by the scope of data, they provide valuable insights into understanding and potential areas for improvement. The contributions of this paper are listed as follows:

\begin{itemize}
        \item \textbf{Emotion-aware CA prototype:} We developed a CA prototype capable of expressing five distinct emotional states using male and female voices. This system provides a controlled platform for investigating human-CA interactions in emotionally charged scenarios.
        \item \textbf{Empirical exploration of user trait-based interaction strategies:} We systematically examined how users with diverse traits (gender, personality, and cultural background) strategize their interactions with emotion-expressive CAs (neutral, happy, sad, angry, fear), uncovering distinctive behavioural patterns in how they interpret and respond to the CAs' emotions.
        \item \textbf{Design implications for future emotion-aware CAs:} Building on our empirical findings, we derived concrete design implications for tailoring emotion-aware CAs to diverse user profiles, demonstrating how accounting for user traits can enable more appropriate and effective emotionally intelligent interactions.
\end{itemize}


\textcolor{black}{The organization of the paper is as follows: Chapter ~\ref{rw} reviews the current research landscape on emotion-aware CAs in relation to user traits and identifies the research gap. Chapter ~\ref{rede} outlines the research pipeline of this study, including the design of emotional scenarios and CA prototyping. Chapter ~\ref{exp} describes the experimental setup, detailing the formulation of two separate studies involving male- and female-voiced CAs and the data collection process. Quantitative and qualitative results are presented in Chapters ~\ref{quan} and ~\ref{qua}, respectively. Key findings are summarized in Chapter ~\ref{find}, followed by a comprehensive discussion in Chapter ~\ref{dis}. Finally, Chapter ~\ref{conc} concludes the research.}

%% file: Sections/Related_Work.tex
\section{Related Work}
\label{rw}

\subsection{Interaction with Conversational Agents}
CAs have become an integral part of daily life, facilitating interactions in a wide array of contexts, including smart homes, healthcare, education, and entertainment~\cite{mctear2017rise}. Previous research has shown that these agents improve user experience by providing natural language interfaces that allow users to interact through voice or text in ways that feel intuitive and efficient~\cite{hoy2018alexa}. Studies focusing on the use of smart speakers, for example, reveal that users engage not only for functional tasks, such as playing music or setting reminders, but also for more social and playful interactions, such as asking for jokes or stories~\cite{porcheron2018voice}. These findings suggest that users value conversational agents not only for their utility, but also for their potential to provide social and emotional engagement.
Despite the rapid proliferation of CAs, much of the research in this area has focused on improving the technical capabilities of these systems, such as natural language processing (NLP) and dialogue management, rather than the emotional dimension of user-agent interactions~\cite{higashinaka2014evaluating}. The emotional capacity of CAs remains underdeveloped, limiting their ability to engage with users in a way that mimics human conversational patterns, which often involve subtle emotional cues and responses~\cite{clavel2015sentiment}. Our study builds on this body of work by exploring how emotional awareness in CAs can enhance the overall quality of user interaction, particularly in emotionally charged scenarios.

\subsection{Emotion-Aware Conversational Agents}
Emotion recognition and response systems have garnered increasing attention in recent years as researchers strive to make CAs more empathetic and emotionally intelligent. Integrating emotional intelligence into CAs involves recognising, interpreting, and responding to users' emotional states, thereby making interactions feel more natural and human-like~\cite{picard1997ective}. Emotion-aware CAs leverage technologies such as sentiment analysis, voice tone recognition, and facial expression detection to infer the user’s emotional state and adjust their responses accordingly~\cite{zeng2009eigenvalue}. However, many current systems are limited to detecting basic emotions—such as happiness, sadness, and anger—without accounting for the complexity or context of the user’s emotional state ~\cite{banziger2012introducing}. Research has shown that simple empathetic responses from agents can improve user satisfaction and trust, particularly in customer service settings where empathy is crucial for resolving issues~\cite{prendinger2005empathic}. However, designing agents that can consistently interpret emotional cues and generate appropriate responses in real-time is a major challenge. Studies indicate that users may become frustrated or disengaged if the emotional response from a CA feels inappropriate, inauthentic, or overly scripted~\cite{edwards2016engaged}. This highlights the delicate balance required between detecting emotions accurately and responding in a way that enhances rather than detracts from the user experience.

\subsection{User Traits and Emotional Interaction with CAs}
Research has increasingly emphasized the role of individual differences in shaping user experiences with technology. Studies in HCI have long recognized that gender, personality, and cultural background can significantly influence how users interact with digital systems \cite{nass2005wired}. For example, gender differences have been observed in communication patterns, with women generally exhibiting more emotionally expressive and empathetic communication styles, while men tend to prioritize problem-solving and efficiency in conversations~\cite{tannen1990gender}. These patterns have been found to extend to human-agent interactions, where women are more likely to seek emotionally supportive responses from CAs, while men prefer more functional or task-oriented exchanges~\cite{reeves1996media}.
Personality traits, particularly as defined by the Big Five model (extraversion, agreeableness, openness, conscientiousness, and neuroticism), also play a crucial role in shaping user behaviour in interactions with CAs~\cite{mccrae1992introduction}. Extraverts, for instance, are more likely to engage actively with CAs and enjoy emotionally rich exchanges, whereas introverts may prefer more reserved and reflective interactions~\cite{sierros2010durable}. Personality-adaptive interfaces, which adjust interaction styles based on the user’s personality, have been shown to enhance the perceived naturalness of the interaction and improve overall user satisfaction~\cite{bickmore2005social}. However, most CAs today lack the ability to dynamically adapt to user personality traits in real-time, limiting the effectiveness of these systems for certain user groups.
Cultural differences further complicate the design of emotionally responsive CAs. Hofstede’s cultural dimensions theory~\cite{hofstede2001culture} suggests that cultural norms and values influence how emotions are expressed and interpreted, which in turn affects communication styles. For example, in collectivist cultures, such as those in many parts of Asia, individuals tend to prioritize group harmony and may suppress overt emotional displays \cite{markus1991cultural}. In contrast, cultures with higher levels of emotional expressiveness, such as in Latin America, tend to encourage more direct and open emotional communication ~\cite{matsumoto2006culture}. These cultural factors are important when designing CAs that are intended to operate in a global context, as users may have varying expectations about how emotionally responsive an agent should be.

\subsection{Research Gap}
While previous research has advanced the technical aspects of conversational agents, there remains a lack of understanding regarding user engagement in emotionally charged interactions. Most studies emphasize emotion recognition and response generation but often overlook how user traits -- such as gender, personality, and culture -- affect these interactions. Our study bridges this gap by empirically examining how users with diverse traits interact with emotion-aware CAs in specific scenarios. By focusing on user strategies and preferences, we provide insights into designing CAs that align with users’ emotional needs, emphasizing the importance of personalisation based on distinct user traits often neglected in prior research.

%

%% file: Sections/Research_Design.tex
\begin{table}[!t]
\centering
\caption{The most-acknowledged emotional scenarios based on the answers provided by the participants in the pre-study.}
\label{tab:pre-study}
\renewcommand{\arraystretch}{1.1} 
\setlength{\tabcolsep}{6pt} 
\begin{small} 
\begin{tabular}{@{}ll@{}}
\specialrule{1.3pt}{0pt}{0pt} 
\textbf{\arraybackslash Emotion} & \textbf{\arraybackslash Which scenario makes you feel the most?} \\
\midrule
\midrule
\textbf{Neutral} & I put on my shoes before leaving the house. \\
\textbf{Happy} & I am visiting my favorite country/city. \\
\textbf{Sad} & I see children suffering from disease, sickness, or war. \\
\textbf{Angry} & I get betrayed by a close friend or relative. \\
\textbf{Fear} & \begin{tabular}[c]{@{}l@{}}I am walking in the dark in the woods when I stumble upon a dead body. \\ The blood seems fresh and I hear a branch breaking from behind.\end{tabular} \\
\specialrule{1.3pt}{0pt}{0pt} 
\end{tabular}
\end{small}
\end{table}

\section{Research Design}
\label{rede}

\subsection{Emotional Scenarios}
Our study draws upon frameworks such as the work of ~\cite{desmet2018measuring} on assessing emotional responses, ensuring that the selected emotional scenarios were relevant and effective in eliciting distinct and meaningful emotional reactions from users. We conducted an online pre-study \cite{zhang2022initial} to identify the most representative emotional scenarios associated with the five basic emotions examined in our study -- \textit{neutral}, \textit{happy}, \textit{sad}, \textit{angry}, and \textit{fear} \cite{gu2019model,tracy2011four}. These emotions encompass a broad spectrum of fundamental emotional states, which are critical to the development of responsive and adaptive AI systems~\cite{assunccao2022overview}. By focusing on these five emotions, our study seeks to understand how individuals respond to and cope with a wide range of emotional experiences. Additionally, we aim to identify human strategies that can inform the development of future emotion-aware conversational agents. 
To achieve this, we developed a set of presumed scenarios for each emotion designed to evoke the strongest emotional responses. These scenarios were created through a workshop discussion involving four of the authors.

A total of 13 participants took part in this pre-study. Each participant was asked to select the scenario within each emotion category that elicited the strongest emotional response. To ensure that the study focused solely on identifying the most recognized emotional scenarios, all responses were collected anonymously, and no personal information was gathered from the participants. The results of the identified five emotional scenarios are presented in Tab. ~\ref{tab:pre-study}.

\begin{figure*}[!t]
    \centering
    \subfigure[Computer views of interacting with the CA. This example shows the \textit{\textbf{happy}} emotion with the male voice. Up: before the conversation; below: after the conversation.]{
        \includegraphics[width=0.38\linewidth]{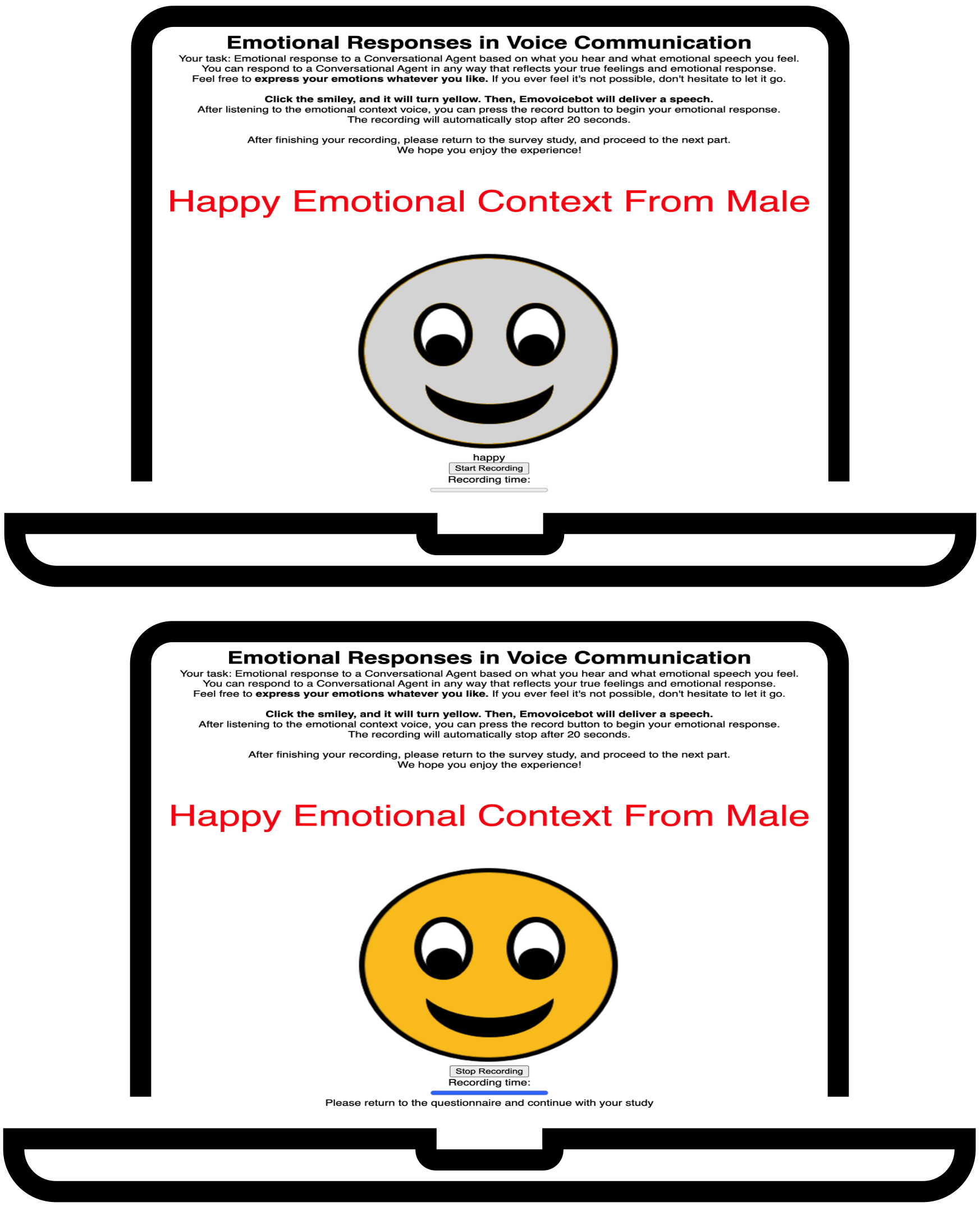}
    }
    \hspace{0.05\linewidth} 
    \subfigure[Mobile views of interacting with the CAs. This example shows the \textit{\textbf{fear}} emotion with the female voice. Left: before the conversation; right: after the conversation.]{
        \includegraphics[width=0.38\linewidth,height=.3\textheight]{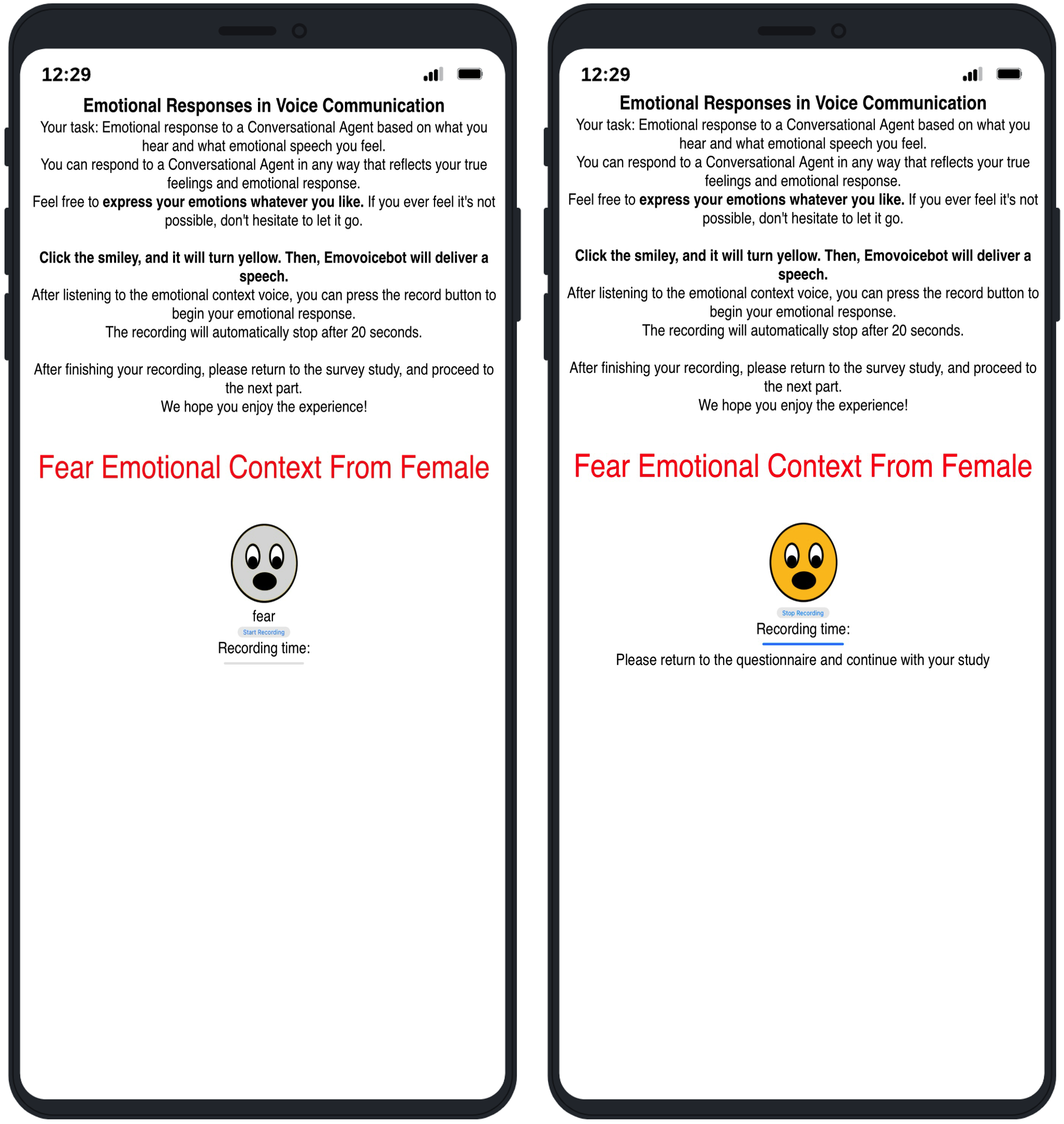}}
    \caption{The interfaces of our CA prototype featuring five emotive smiley icons with each representing a distinct emotion. Users are prompted to click the smiley faces, where the selected icons will turn yellow, which then initiates the conversation phase.}
    \label{fig:ca_views}
\end{figure*}

\subsection{The Conversational Agent}
We developed a prototype of our CA which features a user-friendly interface with five distinct smiley face icons, each representing a core emotion aforementioned. To create emotionally expressive conversational scenarios, we employed the OpenVokaturi\footnote{\url{https://vokaturi.com/news/2022/openvokaturi-4-0}} toolkit to validate the emotional content of the speech samples. OpenVokaturi, which is capable of recognizing five primary emotional states, played a critical role in ensuring emotional consistency across the stimuli used in our study. In addition, it served as the core analytical tool for assessing the emotional characteristics of participants’ spoken responses, offering a robust methodological foundation for advancing research in emotion-aware conversational systems~\cite{ma2025advancing}.
As shown in Fig.~\ref{fig:ca_views}, a uniquely designed smiley face icon was created for representing each distinct emotion to enhance comprehension of the emotional context (all five icons are presented in the Appendix ~\ref{appen.a}).

Users initiate the interaction by selecting one of these icons, which then amplifies and shifts to the center of the screen, setting the stage for an emotionally focused dialogue. This transition highlights the selected emotion and visually prepares the user for a conversation tailored to that emotional context. The CA was designed to support both male and female voices, each articulating \textcolor{black}{the five} emotional scenarios presented in Tab.~\ref{tab:pre-study} while embodying the corresponding emotions to create specific emotion-aware dialogue contexts. The scripted articulations of the five emotional scenarios embedded in our CAs are detailed in the Appendix ~\ref{appen.b}. For each scenario, pre-recorded male and female voice samples, which were ensured for native level, were integrated into the CAs, ensuring that emotional expressions were conveyed effectively through tone, pitch, and speech modulation. For the male voice, as the authors lacked native-level proficiency, a certified voice-changing generator (Murf AI tool\footnote{\url{https://murf.ai/text-to-speech}}) with emotional modulation was used to ensure the utterances were in clear, understandable English. The female voice was recorded by one of the authors, a native English speaker with a multilingual background, providing high-quality, standard English articulation. Prior research~\cite{ma2022should} has highlighted the impact of voice gender on user perceptions. For example, male voices are often perceived as authoritative and neutral, whereas female voices are associated with empathy and warmth~\cite{tolmeijer2021female, mcdonnell2019chatbots, ma2022should}. Therefore, we alternated between male and female voices to explore gender-based differences in user engagement and to minimize potential biases. All audio recordings of CAs were produced at a sample rate of 48 kHz to ensure high-quality output. We conducted multiple iterations of emotional speech generation to ensure their validity, with final selections approved unanimously by all authors and harnessed in our CAs.

We employed single-turn dialogue \cite{lowe2015ubuntu,wan2016match,molino2018cota,chen2019driven,shang2015neural} in our study, where users exchanged one turn each during dialogic interactions. According to Schegloff and Sacks \cite{schegloff1973opening}, even brief exchanges, such as a basic greeting sequence (i.e., “Hi” followed by “Hello”), form a “conversational turn-taking structure.” Similarly, Levinson \cite{levinson1983pragmatics} emphasized that conversations are defined by shared communicative intent. Thus, a single exchange involving the reciprocal flow of information meets fundamental conversational norms. This perspective was reinforced by early chatbot designs like ELIZA \cite{weizenbaum1966eliza}, which engaged users in brief yet meaningful interactions that were recognized as conversations. In our study, the CAs conveyed emotional scenarios while expressing communicative intent. Users in each interaction responded by reciprocally exchanging one turn, following the information thread initiated by the CAs. Therefore, each interaction with our CAs was conferenced as a pragmatic and complete conversation.

In our setup, each single-turn dialogic interaction for each emotional scenario with the CA was designed to allow approximately 20 seconds (in total, 5 emotions*20 seconds) for the user to respond after receiving the CA’s utterance. This decision, made after thorough discussions among the authors, was informed by a review of similar research \cite{ma2023emotion,ma2022should,marcus2017measuring} that utilized a 5-second response time. By extending the response duration, we enabled richer emotional expression and greater context within the dialogue, providing participants with a more immersive and realistic interaction experience. This approach also aimed to improve the reliability of user evaluations by fostering deeper emotional engagement compared to shorter response times.

Once in the conversation phase, the CA initializes according to the chosen emotion, engaging the user in a dialogue that reflects the emotional state represented by the smiley. This design encourages users to experience and respond to varied emotional scenarios, making the interaction feel more natural. By directly involving users in these emotional dialogues, the prototype explores how emotions can shape conversational dynamics, allowing us to examine user strategies in responding to different emotional contexts. This approach not only enhances the emotional realism of the dialogue but also provides valuable insights into user behavior and preferences in emotionally charged interactions with conversational agents.


%% file: Sections/Experiment.tex
\section{Experiment Setup}
\label{exp}

\begin{figure*}[!t]
    \centering
    \includegraphics[width=\linewidth]{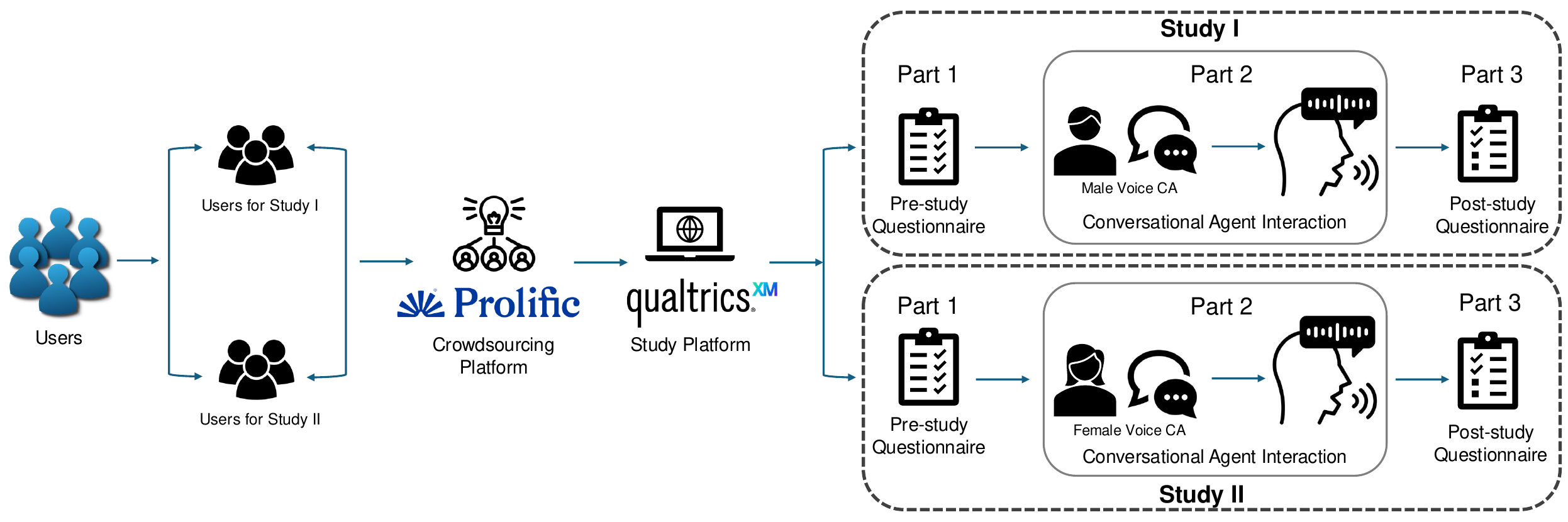}
    \caption{An overview of the user study workflow.}
    \label{fig:us_process}
\end{figure*}

To explore user interactions with the developed emotion-aware CAs regarding user traits -- gender, personality, and cultural background (ethnicity and geographic background), we conducted a controlled crowdsourcing experiment consisting of two studies (Study I and Study II \textcolor{black}{with different participants}).
The primary distinction between the studies was the voice used by the CAs: Study I featured \textcolor{black}{male-voiced CAs, while Study II embraced female-voice CAs}. Both studies employed a within-subjects design, with the independent variables being the five distinct CAs, each expressing a different emotional state.
\textcolor{black}{This setup also ensured a consistent focus on a single voice gender within each study, reducing confounding effects that might arise from mixed-gender voice presentations in a single experimental session.} 

\subsection{Participants}
Participants were recruited through Prolific \cite{palan2018prolific} \footnote{\url{https://www.prolific.com/}}, a platform designed specifically for academic research and well-suited for social and economic science experiments \cite{peer2017beyond}. Unlike other crowdsourcing platforms that focus on microtasks, Prolific offers a diverse participant pool \textcolor{black}{ensuring a diverse sample in terms of gender, personality, and cultural background}, with over half of the participants holding a bachelor’s degree or higher, making it ideal for studies that require reliable data from a varied sample.

We titled our study “Emotional Conversational Agent Interaction” and targeted English-fluent participants to ensure consistent and accurate responses. Prolific pre-screened participants for English fluency before recruitment. In total, we recruited 60 participants, with 30 allocated to each study (I: $M = 32.29, SD = 9.71$ and II: $M = 27.17, SD = 5.02$). To examine the impact of user traits on conversational strategies in emotionally charged scenarios, we balanced gender distribution by recruiting 15 male and 15 female participants per study through Prolific’s pre-screening filters. An unexpected issue occurred when one participant in Study I exceeded the time limit set by Prolific. The participant communicated with authors anonymously via the platform, explaining the situation. After reviewing the responses, which were deemed valuable and valid by the authors, it was decided to include this participant’s data, bringing the total number of participants in Study I to 31.

Participants were compensated £4.50 upon successful completion of the task. Prolific reported that the average hourly payment was £13.90 for Study I and £10.80 for Study II, which are considered above-average compensation rates on the platform. The average completion times were 21 minutes and 25 minutes for studies I and II. Prior to participation, users received detailed instructions and, after giving their consent, were directed to engage with the CAs and complete the questionnaires. \textcolor{black}{All participants were instructed to complete the study in a quiet and distraction-free environment.}

\subsection{\textcolor{black}{Study Procedure}}
We utilized Qualtrics \footnote{\url{https://www.qualtrics.com/}}, a leading online survey platform, to integrate our emotion-aware CAs with two distinct voice conditions (male for Study I and female for Study II). Qualtrics is widely recognized for its robust features and suitability for complex research designs, making it ideal for our study. We developed two separate surveys corresponding to the two studies. Each survey consisted of three main sections: a pre-questionnaire, an emotional CA interaction phase, and a post-study questionnaire. The pre-questionnaire gathered basic demographic information, \textcolor{black}{which} helped contextualize the participants’ subsequent interactions with the CAs. During the conversational interaction task phase, participants engaged with the CAs, exhibiting different emotional expressions. Following this, the post-study questionnaire employed a 7-point Likert scale to evaluate various metrics related to the user experience alongside open-ended questions that allowed participants to elaborate on their responses. Participants were required to answer questions sequentially during the dialogue task, ensuring that each response was submitted before proceeding to the next question, thereby maintaining a focused and orderly flow. Unique survey URLs were generated and embedded in Prolific, which directed participants to the respective studies. Aside from the differing voices (male in study I and female in study II), the content and layout of the questionnaires remained consistent across both studies to ensure comparability of the results.

\begin{table}[!t]
\caption{The quantified metrics measured in our study. The pre-study metrics were measured in 5-point Likert scale, while post-study metrics were recorded in 7-point Likert scale.}
\label{tab:juji_conversational_skills}
\begin{scriptsize}
\begin{tabular}{p{1.8cm} p{1.8cm} p{11cm}}
\specialrule{1.5pt}{0pt}{0pt} 
\toprule
\textbf{Metric}                & \textbf{Pre-/Post-study}                                                                   & \textbf{Question}                                                                         \\ 
\midrule
\textbf{Personality}        & Pre-study & \textbullet\ On a scale of 1 to 5, how would you rate your personality? I see myself as someone who (1 = Strongly Disagree, 5 = Strongly Agree): \newline - is talkative
\newline - is reserved
\newline - is full of energy
\newline - generates lots of enthusiasm
\newline - tends to be quite
\newline - has an assertive personality
\newline - is sometimes shy, inhibited
\newline - is outgoing, sociable
 \\ 
\textbf{Baseline Emotional State}    & Pre-study & \textbullet\ On a scale of 1 to 5, how would you rate your current emotional state? (1 = Very Negative, 5 = Very Positive) \newline \textbullet\ In the past week, how often have you felt the following emotions? (1 = Never, 5 = Very Often) : Happiness; Sadness; Anger; Fear; Surprise; Disgust; Calm    \\ 
\textbf{Familiarity with CAs}         & Pre-study   & \textbullet\ How frequently do you use conversational agents (e.g., Siri, Alexa, Google Assistant)? (1 = Never, 5 = Daily) \newline \textbullet\ How comfortable are you using voice assistants? (1 = Extremely uncomfortable, 5 = Extremely comfortable)  \\
\midrule
\midrule
\textbf{Consistence}   & Post-study         & \textbullet\ To what extent do you think the conversational agent's speech is consistent with the five emotions? (1 = Strongly Inconsistent, 7 = Strongly Consistent)        \\ 
\textbf{Convenience} & Post-study & \textbullet\ Did you find it convenient to interact with the conversational agent with the five different emotions? (1 = Strongly Disagree, 7 = Strongly Agree) \\ 
\textbf{Comfort}      & Post-study     & \textbullet\ How comfortable did you feel when interacting with the conversational agent with the five different emotions? (1 = Extremely Uncomfortable, 7 = Extremely Comfortable) \\ 
\textbf{Influence}    & Post-study  & \textbullet\ How did the conversational agent’s emotion influence your willingness to continue the conversation in the five emotions? (1 = Significantly Decreased, 7 = Significantly Increased) \\ 
\textbf{Alignment}    & Post-study & \textbullet\ Did you feel your emotional strategy aligned with the conversational agent’s emotion in each scenario? (1 = Never Aligned, 7 = Always Aligned) \\ 
\textbf{Effectiveness}    & Post-study    & \textbullet\ How effective do you think your chosen strategy was in maintaining a positive interaction with the conversational agent in the five different emotions? (1 = Very Ineffective, 7 = Very effective) \\
\textbf{Learnability}    & Post-study    & \textbullet\ Did interacting with a conversational agent with the five different emotions influence your understanding of how to handle emotions in real-life conversations? (1 = Strongly Disagree, 7 = Strongly Agree) \\
\textbf{Mutual Connection}    & Post-study     & \textbullet\ Did you feel an emotional connection with the conversational agent during any of the interactions in the five different emotions? (1 = Very Disconnected, 7 = Very Connected) \\ 
\textbf{Agent Awareness}    & Post-study     & \textbullet\ If the conversational agents were emotionally aware, which emotion would you prefer them to express during the conversation? (1 = Strongly Undesirable, 7 = Strongly Desirable) \\ 
\bottomrule
\specialrule{1.5pt}{0pt}{0pt} 
\label{tab:metrics}
\end{tabular}
\end{scriptsize}
\end{table}

The study was designed with accessibility, supporting participation via desktop, laptop, or mobile devices (see Fig. ~\ref{fig:ca_views} for device-specific views). To minimize any variability in user experience due to the device used, we followed Qualtrics’ guidelines for mobile optimization, ensuring that the survey design, user interface, and integration of CAs were consistent across all platforms. Pre-distribution testing with end users (n=4) confirmed that the design and functionality were equivalent on both mobile and desktop devices, providing a uniform experience regardless of the device used by participants. This careful setup and rigorous testing ensured that the study maintained high levels of accessibility, reliability, and data integrity, enabling us to accurately assess the influence of user traits on interactions with emotion-aware conversational agents. An overview of the experiment process is illustrated in Fig. ~\ref{fig:us_process}. \textcolor{black}{The sequence of the five CAs, each linked to distinct emotions, was randomized to be presented to each participant after they initiated their own studies. This was achieved using Qualtrics’ randomization function to minimize potential order effects. Each participant was thoroughly briefed on the concept of single-turn dialogue with the CAs, emphasizing that their role was simply to respond to the utterances provided by the CAs through one reciprocal exchange of turn. During the 20-second response period, participants were free to vocally reply to the CAs in any manner to complete the dialogue.}

\subsection{Data Collection}
\textcolor{black}{Each participant successfully completed the study, providing both valid dialogic interactions and survey responses.} We collected a range of quantified metrics through both pre- and post-study questionnaires to assess various aspects of user interaction with the CAs. In the pre-study phase, \textcolor{black}{aside from collecting information of gender, ethnicity, and residence location,} we \textcolor{black}{also} measured participants’ personality traits, \textcolor{black}{recent} baseline emotional states, and familiarity with CAs using a 5-point Likert scale, adapted from established references \cite{volkel2021eliciting,simms2019does,he2017enhancing,zhang2023industrial}. Specifically, we assessed extroversion (\textcolor{black}{personality}) using the most recognized Big Five questionnaire \cite{caprara1993big}, commonly employed for evaluating human personality. The assessment included eight questions (see Tab.~\ref{tab:metrics}), each quantified on a 5-point scale \cite{andriella2021have}. These metrics helped establish a baseline for each participant’s characteristics prior to interacting with the CAs. In the post-study phase, \textcolor{black}{both quantitative and qualitative measurements were included.} The quantitative metrics focused on evaluating the overall perceived performance and functionality of the CAs, including participants’ impressions of the agents’ responses and the  relationship dynamics between the participants and the CAs. To enhance the precision of these assessments, we employed a 7-point Likert scale to measure nine distinct metrics quantitatively. \textcolor{black}{We harnessed nine key metrics to evaluate the overall functionality and performance of our CAs inspired by previous literature: consistence \cite{yang2019understanding}, convenience \cite{oh2020differences,kim2019comparing}, comfort \cite{yang2019understanding,bickmore2007practical}, influence \cite{ling2021factors,diederich2022design}, alignment \cite{langevin2021heuristic}, effectiveness \cite{yang2021designing,wambsganss2020conversational,jabir2023evaluating}, learnability \cite{winkler2020sara,langevin2021heuristic}, mutual connection \cite{clark2019makes,christoforakos2021connect,luger2016like}, and agent awareness \cite{ma2023emotion,divekar2019you}.} Detailed descriptions of these metrics are provided in Tab. ~\ref{tab:metrics}. Participants proceeded to the post-study survey only after interacting with all five randomized emotional CAs. Notably, we ensured each emotion of every CA was measured across all metrics for every participant (in total, nine subjective questions were asked corresponding to the nine metrics). We think collecting feedback after the full study minimized novelty effects and facilitated comparisons of participants. In addition, we believe filling the survey after each interaction (9*5=45 questions total) risked causing fatigue and tediousness, potentially compromising result validity. Besides, several open-ended responses were collected to provide qualitative insights into user strategies and perceptions. We investigated \textbf{adjustment of responses}, \textbf{engaged strategies}, and \textbf{potential strategic modifications} that will be exhaustively presented in the following parts. The qualitative analysis process was implemented utilizing thematic analysis, while the integration of quantitative and qualitative methods offered a holistic understanding of user interactions with emotion-aware CAs.

%% file: Sections/Quantitative_results.tex
\section{Quantitative Results}
\label{quan}

We first present our participants' demographic and background information \textcolor{black}{of the two studies}, followed by detailed \textcolor{black}{quantitative} results from \textcolor{black}{Study I and Study II separately}.



\begin{figure*}[!t]
    \centering
    \includegraphics[width=\linewidth]{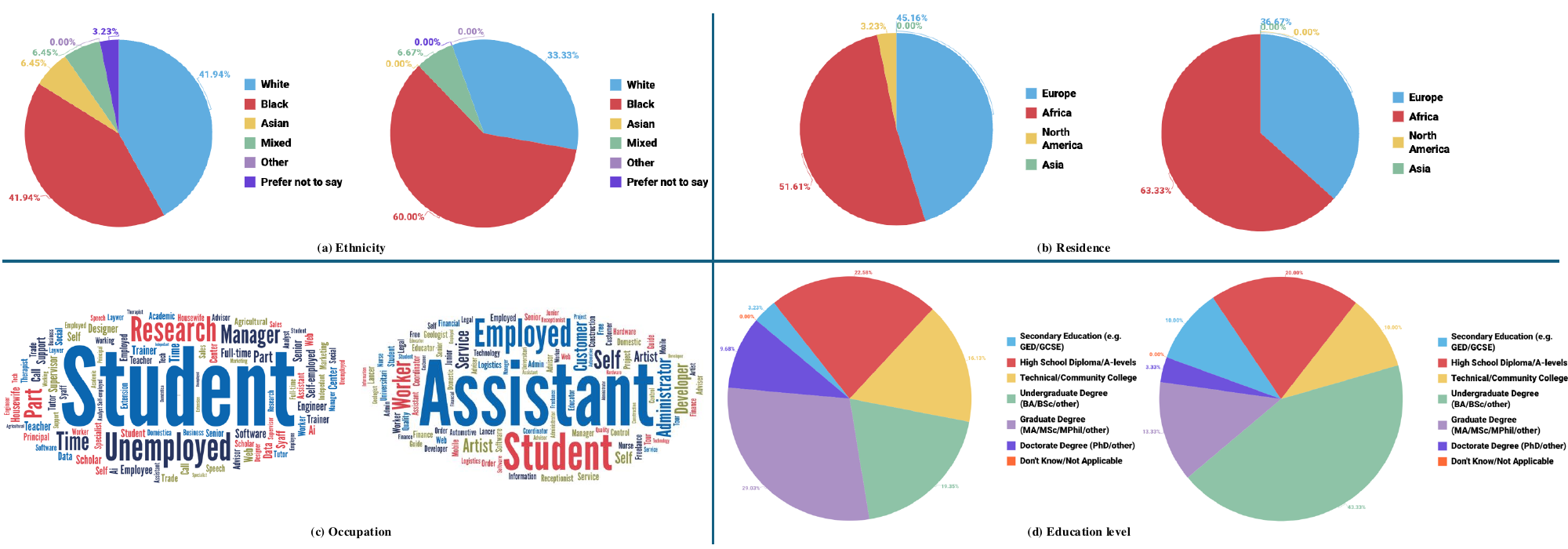}
    \caption{Some of the demographic information about the participants. In subfigures: \textbf{\textit{Left: Study I; right: Study II.}}}
    \label{fig:demographics}
\end{figure*}

\subsection{General Demographics and Background}
We received a total of 61 completed responses. We pre-screened the participants based \textcolor{black}{on} an evenly-distributed assignment in terms of sex (Male, Female) \cite{geary1998male}. However, we still asked them to provide information about their genders \cite{todd2019demographic} (Man (including Trans Male/Trans Man), Woman (including Trans Female/Trans Woman), Non-binary), and ethnicity (White, Black, Asian, Mixed, Other) (Fig. ~\ref{fig:demographics}.$a$) based on the categorization of Prolific. As a result, gender distribution was balanced across both studies (Study I: 16 Men and 15 Women; Study II: 15 Men and 15 Women). Additionally, we gathered data on the LGBTQ+ status of participants \cite{taylor2024cruising}, where 12.9\% from Study I and 13.33\% from Study II reported inclusion of the LGBTQ+ community. Fig. ~\ref{fig:demographics}.$b$ illustrates the geographic distribution of participants by continent of residence. Fig. ~\ref{fig:demographics}.$c$ and ~\ref{fig:demographics}.$d$ depict participants’ occupation and education level, respectively. This highlights the diversity of occupations within our study sample and, importantly, shows that nearly 60\% of participants in each study hold an undergraduate degree or higher. \textcolor{black}{In terms of personality by using the Big Five model, we found that in Study I: 9 participants identified as introverts, 9 as extraverts, and 13 as ambiverts; as in Study II: 13 participants were categorized as introverts, 8 as extraverts, and 9 as ambiverts.}





\begin{figure*}[!t]
    \centering
    \includegraphics[width=\linewidth]{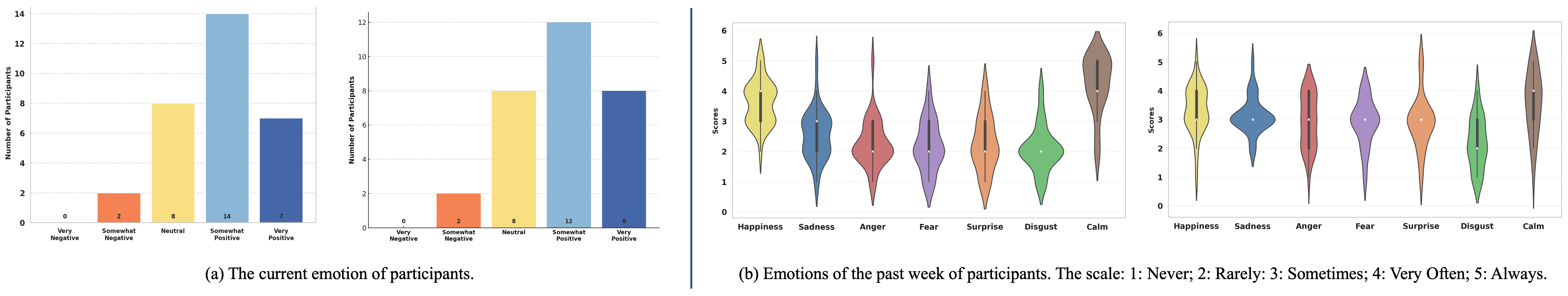}
    \caption{The baseline emotions of participants. \textbf{\textit{Left: Study I; right: Study II.}}}
    \label{fig:baseline_emo}
\end{figure*}

Furthermore, we assessed the baseline emotional states of participants. Figure~\ref{fig:baseline_emo}.$a$ shows that the majority of participants were in a positive emotional state at the time of assessment in both studies. Regarding their recent emotional states, most participants reported positive emotions, such as happiness and calmness, as depicted in Figure~\ref{fig:baseline_emo}.$b$. Based on these results, we can confidently state that participants were largely in stable emotional states prior to and during the study, which likely helped mitigate unnecessary psychological biases and errors.

In Study I, participants’ familiarity with CAs was notably high, with 45\% and 23\% reporting daily and weekly usage, respectively. Google Assistant emerged as the most popular CA, used by 77\% of participants, followed by Alexa and Siri, each at 39\%. Regarding comfort levels, 39\% of users felt somewhat comfortable interacting with CAs, while 33\% reported feeling extremely comfortable. In Study II, 30\% and 33\% of participants reported daily and weekly usage of CAs, respectively. Google Assistant remained the most frequently used CA at 77\%, followed by Siri at 57\% and Alexa at 33\%. Comfort with CA interaction was also similar, with 37\% of participants feeling somewhat comfortable and 33\% feeling extremely comfortable. Our participant pools in both studies demonstrate a strong association with the use of CAs.



\subsection{Results from Study I: Male-Voiced CA}
In this part, we present results from Study I, including the post-study quantitative parts regarding the nine quantified metrics which address RQ1, same applied to that in Study II.

\begin{figure*}[!t]
    \centering
    \includegraphics[width=\linewidth]{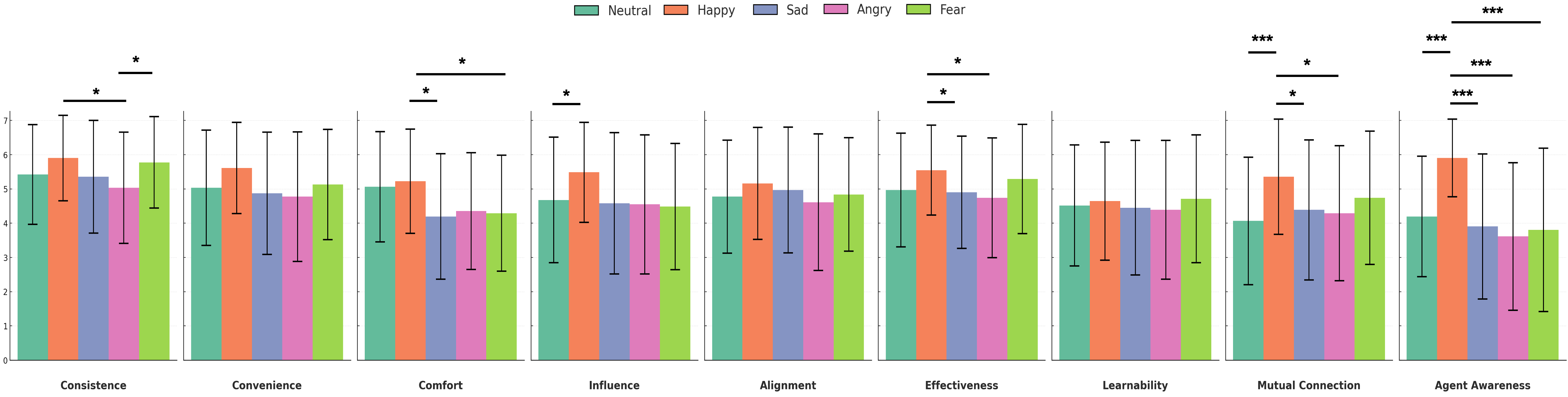}
    \caption{Quantitative results from \textbf{\textit{Study I with male voice}} in CAs, showing the mean ratings across the nine metrics. Pairwise significance from post-hoc analysis is indicated as follows: $*$: $p < .05$; $***$: $p < .001$.}
    \label{fig:quan_I}
\end{figure*}

We asked participants to evaluate the overall performance of the CA prototype and their perceptions during interactions with the CAs. The results of these measurements are presented in Fig. ~\ref{fig:quan_I}. Prior to analysis, normality tests were conducted for each metric, confirming that all metrics followed a normal distribution. Consequently, a one-way repeated measures (rm) ANOVA was performed on all metrics to determine significant differences, followed by post-hoc analysis when significant differences were identified.

\begin{itemize}
    \item \textbf{\textit{Consistence:}} Participants perceived that the CAs' speech was generally consistent with the assumed emotions, especially for \textit{happy} and \textit{fear}, which were rated as the most consistent. The rm ANOVA indicated that ratings differed significantly among the five emotions ($F(3.295, 98.858) = 3.479, p < .05$). Post-hoc analysis with Bonferroni adjustment showed significant differences between \textit{happy--angry} and \textit{angry--fear}.

    \item \textbf{\textit{Convenience:}} Participants found interacting with a CA that expressed a \textit{happy} emotion most convenient, followed by the \textit{fear} emotion, whereas the \textit{angry} emotion was the least preferred. Despite these trends, the statistics revealed no significant differences in convenience ratings across emotions ($F(4, 120) = 2.084, p = 0.088$).

    \item \textbf{\textit{Comfort:}} Participants reported the highest comfort levels when interacting in a \textit{happy} context, followed by \textit{neutral}. Negative emotions, particularly \textit{sad}, received the lowest comfort ratings. Significant differences were found among the emotions ($F(4, 120) = 5.321, p < .001$). Post-hoc analysis revealed significant differences between \textit{happy--sad} and \textit{happy--fear}.

    \item \textbf{\textit{Influence:}} The \textit{happy} context most increased participants' willingness to continue the conversation, showing a clear lead over other emotions, which were rated similarly. The rm ANOVA found significant differences among the conditions ($F(3.171, 95.144) = 2.732, p < .05$). Post-hoc analysis showed a significance between \textit{happy--neutral}.

    \item \textbf{\textit{Alignment:}} Participants felt that emotional strategies were generally aligned with all CAs' emotions, with \textit{happy} being the most aligned and \textit{angry} the least. However, no significant differences identified in alignment ratings ($F(4, 120) = 0.824, p = 0.513$).

    \item \textbf{\textit{Effectiveness:}} Participants rated their strategies for maintaining a positive conversation as effective across all five emotions, particularly for the \textit{happy} CA, followed closely by \textit{fear}. The measurement revealed significant differences among the emotions ($F(4, 120) = 2.894, p < .05$) while post-hoc showed significant differences between \textit{happy--sad} and \textit{happy--angry}.

    \item \textbf{\textit{Learnability:}} For potential generalization, although participants generally did not perceive a high level of learnability in real-life application from interacting with CAs, \textit{fear} was seen as the most beneficial for learning, whereas \textit{angry} ranked the lowest. No significant differences were disclosed in learnability ratings ($F(4, 120) = 0.847, p = 0.498$).

    \item \textbf{\textit{Mutual Connection:}} Participants reported the strongest emotional connection with \textit{happy} CAs, followed by \textit{fear}, while \textit{neutral} was perceived as the least connected. We identified significant differences among the emotions ($F(4, 120) = 5.893, p < .001$). Notably, significant pairwise differences existed between \textit{happy--neutral}, \textit{happy--sad}, and \textit{happy--angry}.

    \item \textbf{\textit{Agent Awareness:}} When asked about the most desirable emotion for emotion-aware CAs, participants overwhelmingly preferred the \textit{happy} emotion, with \textit{neutral} being the second most desired. Negative emotions received much lower ratings. Significance was confirmed: $F(4, 120) = 13.091, p < .001$. Post-hoc analysis revealed significant differences between \textit{happy} and all other emotions.
\end{itemize}

\subsection{Results from Study II: Female-Voiced CA}
\begin{figure*}[!t]
    \centering
    \includegraphics[width=\linewidth,height=4cm]{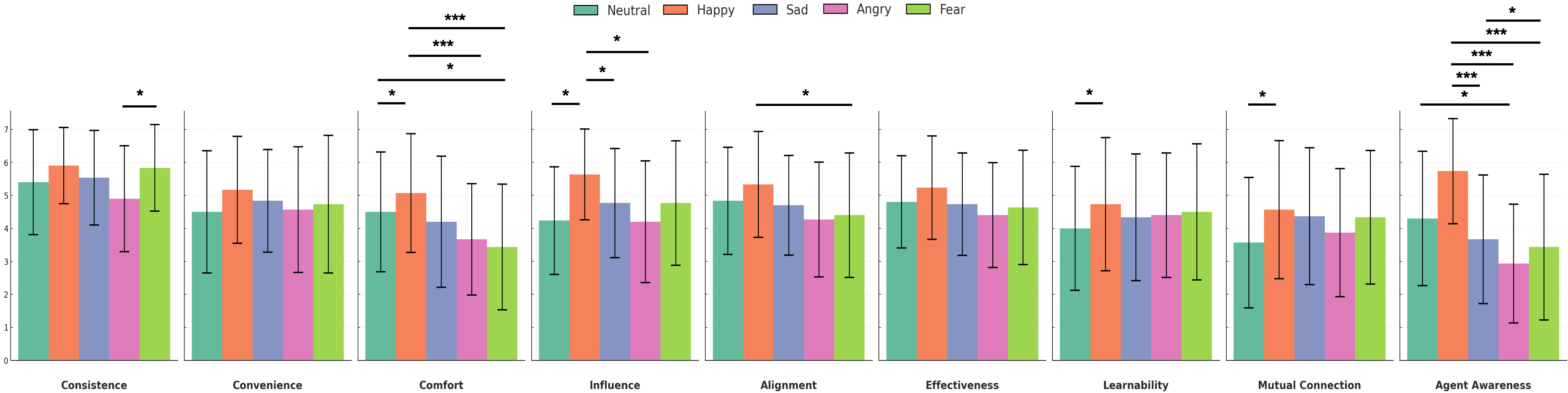}
    \caption{Quantitative results from \textbf{\textit{Study II with female voice}} in CAs, showing the mean ratings across the nine metrics. Pairwise significance from post-hoc analysis is indicated as follows: $*$: $p < .05$; $***$: $p < .001$.}
    \label{fig:quan_II}
\end{figure*}

We now present results from Study II, which were quantified and subjected to statistical analysis, as displayed in Fig. ~\ref{fig:quan_II}.

\begin{itemize}
    \item \textbf{\textit{Consistence:}} Similar to results from study I, participants in Study II perceived the CAs' speech as generally consistent with the intended emotions, particularly \textit{happy} and \textit{fear}, which were rated highest, while \textit{angry} was rated lowest. The statistical analysis indicated significance among the five emotions ($F(3.387, 116) = 3.411, p < .05$). Post-hoc showed significance between \textit{angry--fear}.

    \item \textbf{\textit{Convenience:}} Participants found CAs expressing a \textit{happy} emotion most convenient to interact with, followed by \textit{sad}, whereas \textit{angry} was least preferred. Despite these trends, no significant differences in convenience ratings were found across emotions ($F(4, 116) = 1.407, p = 0.236$).

    \item \textbf{\textit{Comfort:}} Participants reported the highest comfort when interacting with CAs expressing \textit{happy}. Negative emotions, particularly \textit{angry} and \textit{fear}, were associated with the lowest comfort ratings. The significant differences were revealed among the emotions ($F(4, 120) = 12.044, p < .001$). with pairwise significance between \textit{neutral--happy}, \textit{neutral--fear}, \textit{happy--angry}, and \textit{happy--fear}.

    \item \textbf{\textit{Influence:}} The \textit{happy} context most increased participants' willingness to continue the conversation, considerably more than the other emotions, with \textit{angry} being the least influential ($F(3.686, 106.904) = 6.952, p < .001$). Post-hoc found significance between \textit{happy--neutral}, \textit{happy--sad}, and \textit{happy--angry}.

    \item \textbf{\textit{Alignment:}} Similar to study I, participants felt that emotional strategies were generally aligned with CAs' emotions, with \textit{happy} rated as the most aligned and \textit{angry} as the least. The rm ANOVA found significant differences in alignment ratings ($F(4, 116) = 4.042, p < .05$). Post-hoc analysis indicated significance between \textit{happy--fear}.

    \item \textbf{\textit{Effectiveness:}} Participants rated their strategies for maintaining a positive conversation as effective across all emotions, especially with the \textit{happy} CA, followed by \textit{neutral}. The \textit{angry} emotion was rated least effective. Significance was disclosed: $F(4, 116) = 2.952, p < .05$, but post-hoc analysis did not reveal any pairwise significance.

    \item \textbf{\textit{Learnability:}} Participants generally did not perceive high learnability for real-life application from interactions with CAs, though \textit{happy} was seen as most beneficial and \textit{neutral} was rated lowest, which differs from Study I. Through analysis, we found significant differences in learnability ratings ($F(3.010, 87.295) = 3.455, p < .05$). Post-hoc analysis indicated a significant difference between \textit{happy--neutral}.

    \item \textbf{\textit{Mutual Connection:}} Participants felt the strongest emotional connection with \textit{happy} CAs, followed by \textit{sad} and \textit{fear}, while \textit{neutral} was perceived as the least connected. Significance among the emotions: $F(3.503, 101.588) = 4.562, p < .005$; Post-hoc pair wise significance: \textit{happy--neutral}.

    \item \textbf{\textit{Agent Awareness:}} Participants preferred \textit{happy} as the most desirable emotion for emotion-aware CAs, with \textit{neutral} as the second choice. Negative emotions, especially \textit{angry}, were least desired ($F(4, 116) = 16.759, p < .001$). Post-hoc analysis identified significance between \textit{neutral--angry}, \textit{happy--sad}, \textit{happy--angry}, \textit{happy--fear}, and \textit{sad--angry}.
\end{itemize}

%% file: Sections/Qualitative_Results.tex
\section{Qualitative Results}
\label{qua}

In this section, we present the qualitative results targeting RQ2, that investigated the relations between user traits and the conversational strategies users used, including three \textcolor{black}{qualified} metrics to thoroughly investigate the interaction strategies through the contextual dialogues -- {\textbf{Adjustments of Responses, Engaged Strategies, and Potential Modifications} evaluated across user traits -- gender, personality, ethnicity, and geographic background. Due to the limited subject size and characteristics, we only analyzed "Black and White" in terms of ethnicity and "European and African" in terms of geographic background in some metrics. We also \textcolor{black}{distilled the summarization of our results and tabulated them} in Fig. ~\ref{fig:qua_I} for Study I and Fig. ~\ref{fig:qua_II} for Study II for better readability.

\subsection{\textcolor{black}{Qualitative Analysis Process}}
We employed a thematic analysis supported by a LLM to enhance the efficiency and depth of the qualitative process \cite{zhang2023redefining,de2024performing}. This method combined the flexibility and structure of thematic analysis with the computational capabilities of LLMs, allowing a systematic and nuanced exploration of participant responses \cite{braun2006using}. Qualitative data were collected through open-ended questions in the post-study survey, where participants detailed their \textbf{Adjustments of Responses, Engaged Strategies, and Potential Modifications} through their interactions with CAs.


We primarily utilized a ChatGPT 4o-based LLM to support our qualitative analysis. Four authors manually reviewed and tabulated the data to gain a comprehensive understanding of participant responses and actively participated in discussions. ChatGPT was then employed to identify recurring phrases, concepts, and key patterns, generating machine-assisted codes that were cross-validated by the authors to ensure alignment with the study’s objectives and capture context-specific nuances. Building on this foundation, ChatGPT  further assisted in grouping related codes into broader categories, which were iteratively refined into cohesive and meaningful themes.

To ensure the validity and reliability of the LLM-generated results, the themes identified by ChatGPT were compared with manual coding independently performed by the four authors. This dual approach minimized bias and ensured the themes were both data-driven and contextually accurate. Following detailed comparisons and discussions among the four authors, the refined themes (presented in sections below) -- spanning each user trait (\textbf{\textit{gender}}, \textbf{\textit{personality}}, \textbf{\textit{ethnicity}}, and \textbf{\textit{geographic background}}) across all three qualitative metrics -- were formulated and then collaboratively reviewed by all authors to ensure their relevance and alignment with design principles. This rigorous process seamlessly integrated the efficiency of LLM-driven analysis with the critical insights of human expertise. In light of the distinction between male and female voices and the differences in collected data, we conducted two separate qualitative analyzes for Studies I and II, applying the same reasoning mechanisms for consistency.

\subsection{Study I: Male-Voiced CA}
\subsubsection{Adjustment of Responses}
\leavevmode \\ \hspace*{.8em} 
\textcolor{black}{\textbf{\textit{Gender}}: \textbf{Gender affects responses: men mirror emotions; women adapt with empathy and tone matching.}}


When asked how they adjusted their responses to a CA’s emotions, men tended to focus on practicality, often matching the agent’s tone but maintaining neutrality and logic. For example, a man said: \textit{"I responded with the same emotion but kept my tone neutral"}. In contrast, women showed more emotional engagement, adjusting their responses to support the agent emotionally, with one saying: \textit{"I made sure the agent felt understood"}. This reflects a broader pattern where men prioritize logic and efficiency while women emphasize empathy and emotional connection.

\begin{figure*}[!t]
    \centering
    \includegraphics[width=\linewidth]{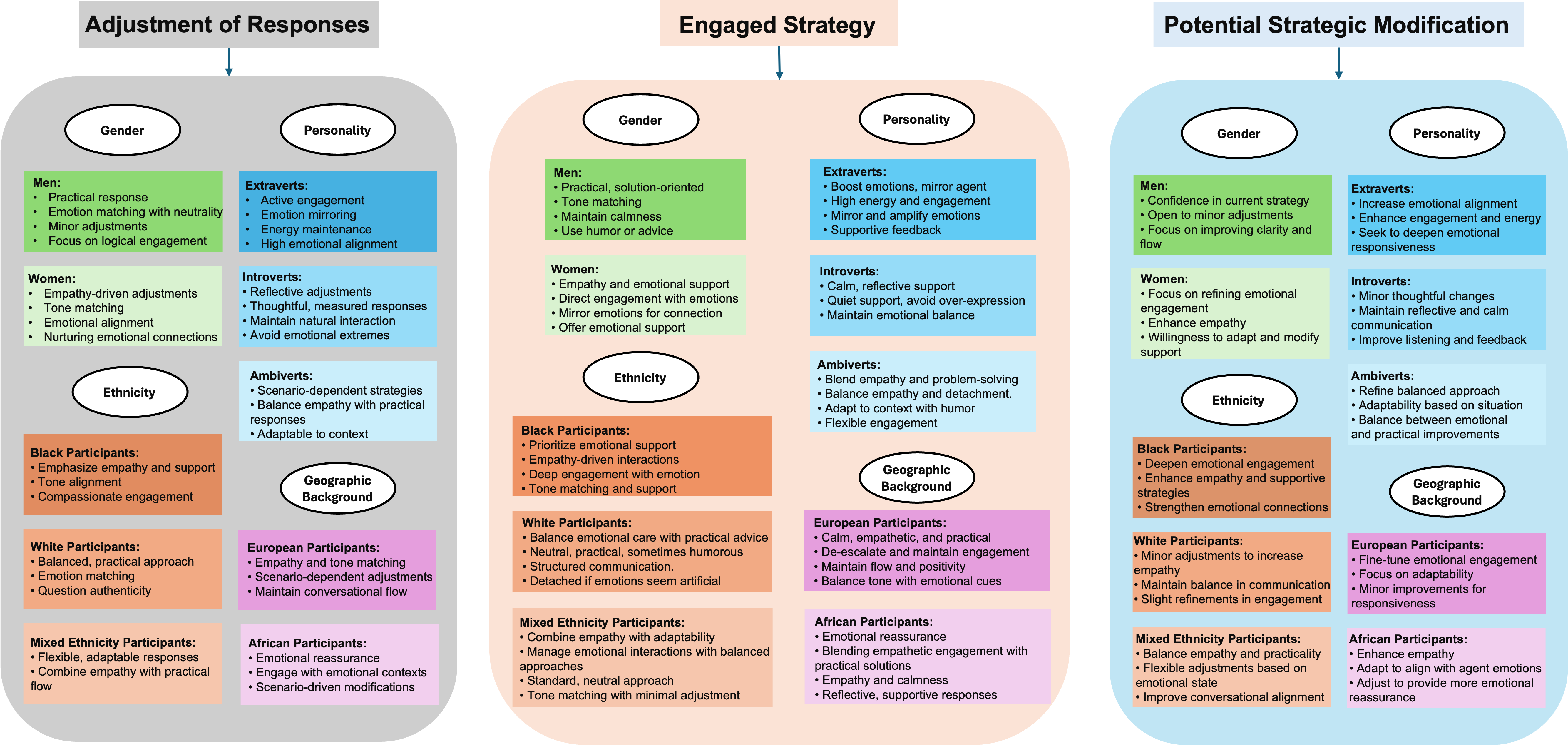}
    \caption{Tabulated summarization of qualitative results from \textbf{\textit{Study I with male voice}} in CAs. The three qualitative metrics are displayed in color-coded rounded rectangles, each assigned a unique color. Inside each rounded rectangle, the four user traits are depicted within white ellipses, while their subcategories with themes are presented below in colored boxes using varying hues.}
    \label{fig:qua_I}
\end{figure*}


\textcolor{black}{\textbf{\textit{Personality}}: \textbf{Personality shapes responses: extraverts engage actively, introverts adjust reflectively, and ambiverts balance empathy with context.}}

Participants adjusted their responses to emotion-aware CAs based on their personality (extraversion, in our paper). Ambiverts displayed a balanced approach, shifting between emotional engagement and neutrality as needed, maintaining flexibility. For example, \textit{"I matched the agent's emotion but kept it balanced"}, said by one. Introverts were more cautious and reflective, focusing on calm, thoughtful responses without becoming too emotionally involved. They preferred to provide composed feedback. Extraverts, by contrast, were highly emotionally engaged, actively mirroring the agent's emotional state, whether upbeat or reassuring. These differences show how ambiverts adapt fluidly, introverts focus on calm reflection, and extraverts thrive on emotional energy in conversations.

\textcolor{black}{\textbf{\textit{Ethnicity}}: \textbf{Ethnicity influences responses: Blacks and Mixed prioritize empathy and tone matching, while Whites balance empathy with scepticism.}}

When adjusting their responses, participants from different ethnic groups showed distinct approaches. Black participants emphasized empathy, adjusting their tone to offer emotional support, especially when the agent expressed sadness or fear. For example, one expressed: \textit{"I adjusted my tone to be more compassionate, making sure the agent felt heard"}. White participants balanced emotional engagement with practicality, matching the agent's tone but maintaining clarity. A typical response was: \textit{"I adjusted to match the agent's emotion but kept the conversation clear"}. Mixed-ethnicity participants were flexible, blending empathy with maintaining conversational flow. One participant reported: \textit{"I tried to be empathetic while keeping the conversation moving"}. These approaches reflect how cultural backgrounds shape communication styles in emotional interactions.

\textcolor{black}{\textbf{\textit{Geographic Background}}: \textbf{Geographical background shapes responses: Europeans emphasize empathy and tone matching, while Africans focus on support and tone alignment.}}

In addition, cultural norms significantly shaped communication styles across different regions. In Europe, participants balanced emotional engagement with practicality, maintaining clarity and efficiency while aligning with the agent’s tone. For example, a participant noted: \textit{"I matched the agent’s tone but kept it practical"}. Asians favored reserved communication, emphasizing harmony and emotional restraint, with responses like: \textit{"I maintained a calm, respectful tone, avoiding strong reactions"}. In African countries, empathy and emotional support were prioritized, especially during negative emotional states, as one Nigerian participant shared: \textit{"I adjusted to show empathy and ensure the agent felt understood"}. From limited samples of Latin America, we found participants were highly expressive, mirroring emotions and offering comfort, as highlighted by a Brazilian saying: \textit{"I matched the emotion and tried to improve the mood"}.

\subsubsection{Engaged Strategies}
\leavevmode \\ \hspace*{.8em} 
\textcolor{black}{\textbf{\textit{Gender}}: \textbf{Gender shapes interaction strategies: men prioritize practical solutions, tone matching, and calmness, while women emphasize empathy, support, and direct engagement with emotions.}}

Men and women show distinct strategies in how they engage with emotions with CAs. Men often focus on practicality and logic, keeping a neutral tone. For instance, in \textit{happy} situations, a participant reported: \textit{"I kept the conversation upbeat but didn’t overly engage with the emotion"}. When faced with \textit{angry} or \textit{fear}, men are more likely to offer solutions or advice, aiming to de-escalate the situation calmly. Women, however, are more likely to adjust their responses to reflect emotional alignment and empathy. In positive moments, for example, \textit{"I matched the agent’s enthusiasm and kept the conversation lively"}, expressed by a woman. In situations of sadness or \textit{fear}, women often offer reassurance, aiming to make the agent feel supported. Their approach emphasizes emotional rapport and comfort.

\textcolor{black}{\textbf{\textit{Personality}}: \textbf{Personality shapes strategies: extraverts boost engagement, introverts provide calm support, and ambiverts mix empathy with practical solutions, adapting to context.}}

Personality, another key user trait, significantly influences the development of effective conversational strategies. Ambiverts balance engagement and practicality, adapting fluidly. In \textit{neutral} or \textit{happy} settings, they align with the agent’s mood but maintain control, stating: \textit{"I tried to match the mood but didn’t let it control the conversation"}. When emotions intensify, they blend empathy with a practical focus. Introverts approach with caution, offering calm and thoughtful responses. In \textit{neutral} interactions, they keep a reserved tone: \textit{"I listened carefully and responded thoughtfully without getting too emotional"}. In intense situations, they provide quiet reassurance without becoming overwhelmed. Extraverts are highly engaged, mirroring the agent’s emotions dynamically. In positive scenarios, one responded: \textit{"I matched the agent’s excitement and kept the conversation energetic"}. During negative interactions, extraverts offer active reassurance, thriving on emotional connection and responsiveness.

\textcolor{black}{\textbf{\textit{Ethnicity}}: \textbf{Ethnicity influences strategies: Blacks prioritize empathy, Whites balance support and advice, Mixed blend empathy and problem-solving.}}

Ethnic differences also impact user strategies in the context of emotional CAs. Black participants prioritized empathy and support, particularly in situations involving sadness or \textit{fear}, with one saying: \textit{"I adjusted my tone to be compassionate and ensure the agent felt supported"}, reflecting a community-oriented approach. In positive scenarios, they mirrored the agent’s joy, maintaining engaging dialogue. White participants balanced emotional engagement with structure, responding positively in \textit{neutral} or \textit{happy} contexts but keeping it clear and practical, as one noted: \textit{"I matched the agent’s tone but kept it straightforward"}. In tense moments, they leaned towards problem-solving, blending advice with emotional responsiveness. Mixed-ethnicity participants demonstrated adaptability, aligning their responses with the agent’s emotional state. In \textit{happy} moments, they matched enthusiasm, while in challenging scenarios, they combined empathy with practical advice, as expressed by one: \textit{"I showed empathy but also offered practical advice to help the agent"}.

\textcolor{black}{\textbf{\textit{Geographic Background}}: \textbf{Residence alters: Europeans balance calmness, empathy, and advice; Africans focus on reassurance, support, and empathetic solutions.}}

Participants from different countries adjust their responses according to cultural communication norms. Participants from Europe balanced emotional engagement with clarity. In \textit{happy} interactions, they matched the agent’s enthusiasm while keeping the conversation practical, and in more challenging situations, like \textit{angry} or \textit{fear}, they focus on de-escalating emotions through logical solutions and calm advice. Asian participants prioritized emotional restraint and harmony. Regardless of the agent's emotions, they responded with controlled, polite, and measured tones. In \textit{happy} moments, they may acknowledge the agent’s joy but remain calm, and in emotional situations like \textit{angry} or \textit{fear}, they offer quiet support, avoiding confrontation to maintain social harmony. In African countries, empathy and emotional connection are emphasized. Participants adjust their tone to reflect the agent's emotions, offering deep reassurance and support, especially in cases of sadness or fear. Even in \textit{neutral} situations, they maintain an emotionally connected tone, ensuring the agent feels supported.

\subsubsection{Potential Strategic Modification}
\leavevmode \\ \hspace*{.8em} 
\textcolor{black}{\textbf{\textit{Gender}}: \textbf{Gender affects openness: men are generally confident but consider minor tweaks, while women are more inclined to refine emotional engagement and boost empathy.}}

Regarding potential modifications, gender-based differences were evident in how men and women adjusted their communication strategies for future interactions. Men generally preferred sticking to their original strategies, showing confidence in their initial approach, and valuing structure and clarity. Typical responses included: \textit{"No, I wouldn’t change anything; my approach was effective"}. When men considered adjustments, they focused on enhancing conversational flow or clarity rather than emotional depth, reflecting a practical, efficiency-driven mindset. In contrast, women were more inclined to adapt their strategies to enhance emotional alignment and support the CA. Common responses like: \textit{"Yes, I would modify my strategy to offer more emotional support"}, underscored their emphasis on empathy and nurturing connections. Women frequently reassessed the emotional effectiveness of their responses, showing a willingness to adjust tone, empathy, and engagement to foster better emotional rapport. This suggests that men prioritize consistency and structure, while women favor emotional adaptability, highlighting distinct gender approaches to communication.

\textcolor{black}{\textbf{\textit{Personality}}: \textbf{Personality tweaks strategy: introverts make cautious changes, extraverts boost connection, ambiverts refine balance.}}

When considering modifications to their conversational strategies, participants with different personality dimensions exhibited unique approaches aligned with their natural communication styles. Ambiverts, balancing introversion and extraversion, were open to minor tweaks based on context, reflecting adaptability without major changes, e.g., \textit{"I might tweak my approach depending on the agent’s emotions"}. Introverts, favoring consistency and control, largely felt their initial strategies were effective, with typical feedback like: \textit{"I wouldn’t make any big changes"}. Any suggested adjustments were subtle, focusing on improving listening or offering more thoughtful feedback rather than increasing expressiveness. Extraverts, the most willing to modify their strategies, aimed to enhance emotional engagement, energy, and support in their interactions, often saying: \textit{"I would definitely modify my approach to be more engaging and emotionally responsive"}. These distinct tendencies highlight how personality shapes openness to strategic adjustments in conversational contexts."

\textcolor{black}{\textbf{\textit{Ethnicity}}: \textbf{Ethnicity guides strategy: Whites favor minor adjustments, Blacks deepen engagement, Mixed balance empathy and adaptability.}}

Strategic modifications vary across different ethnic groups, shaped by emotional engagement and cultural norms. Black participants prioritized enhancing emotional connections, especially in charged conversations, with responses like: \textit{"Yes, I would modify my strategy to offer more emotional support and ensure the agent feels understood"}. Highlighting a focus on empathy and community-driven communication. White participants leaned towards balancing emotional engagement with conversational efficiency, often adjusting strategies to improve flow and structure, as seen in one comment: \textit{"I would adjust my approach to make the conversation clearer and more efficient"}, emphasizing practicality over deep emotional ties. Mixed ethnicity participants showcased adaptability, modifying their strategies based on the emotional context, with statements such as: \textit{"I would align my strategy with the agent’s emotions while keeping the conversation productive"}, reflecting their ability to blend empathy and clarity effectively.

\textcolor{black}{\textbf{\textit{Geographic Background}}: \textbf{Residence for adjustments: Europeans fine-tune engagement and flexibility, Africans enhance empathy and adaptability.}}

Participants from different countries displayed nuanced approaches to modifying their conversational strategies, mostly determined by their cultural values and communication norms. Participants from Europe prioritized enhancing efficiency and clarity, aiming for direct and structured conversations. A common sentiment was: \textit{"I would adjust my strategy to make the conversation clearer and more effective"}, highlighting a focus on practicality. Asians tended to maintain harmony and emotional restraint. Participants preferred keeping interactions calm and polite, with responses like \textit{I would adjust to ensure the conversation stayed respectful"}, reflecting their cultural focus on harmony. African participants emphasized empathy and emotional connection, often modifying their strategies to offer greater support and encouragement. One commented: \textit{"I would modify my strategy to provide more emotional support"}, underscoring a cultural emphasis on community and emotional engagement.

\subsection{Study II: Female-Voiced CA}

\subsubsection{Adjustment of Responses}
\leavevmode \\ \hspace*{.8em} 
\textcolor{black}{\textbf{\textit{Gender}}: \textbf{Gender shapes interaction: women prioritize emotional connection, men prefer structured adjustments.}}


\begin{figure*}[!t]
    \centering
    \includegraphics[width=\linewidth]{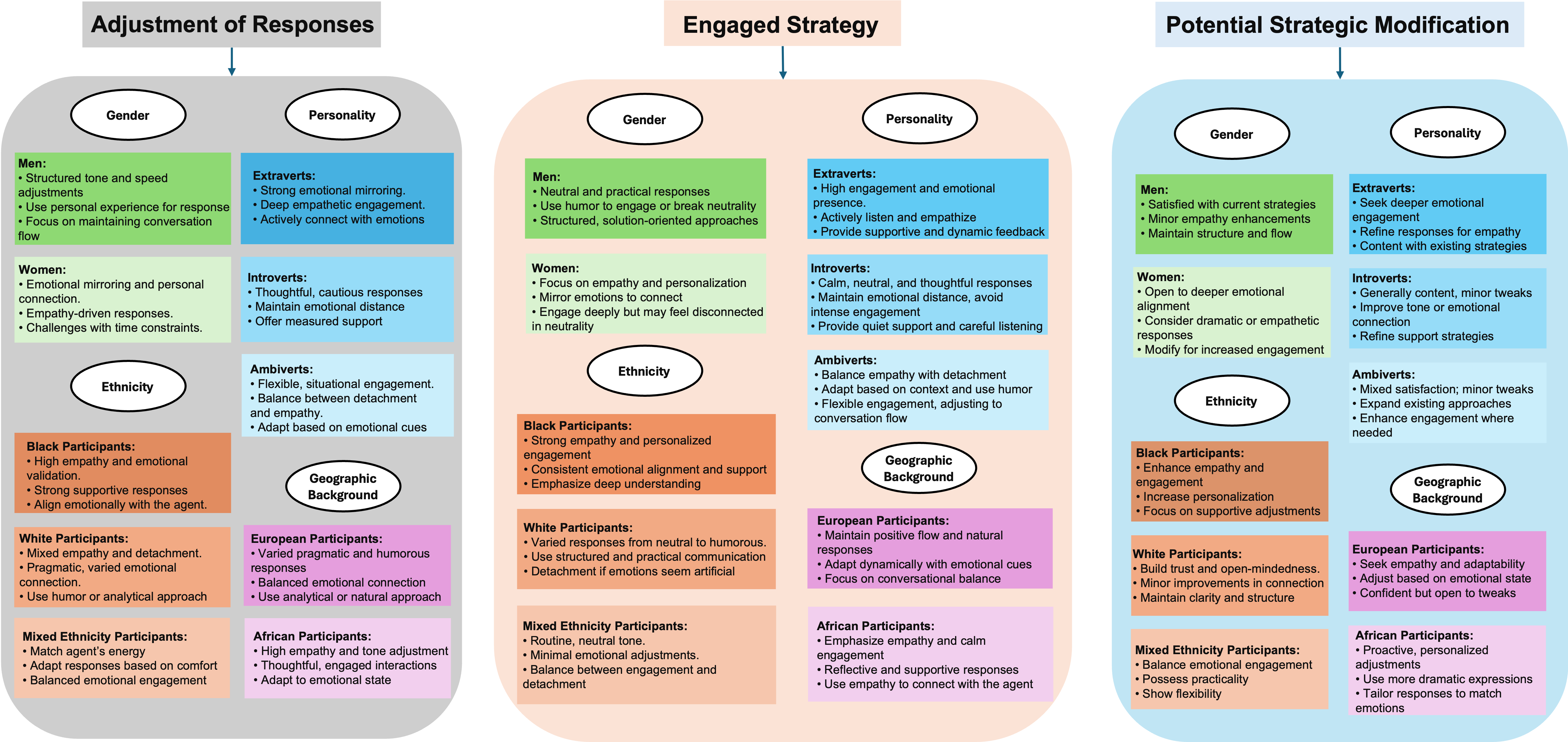}
    \caption{Tabulated summarization of qualitative results from \textbf{\textit{Study II with female voice}} in CAs. The three qualitative metrics are displayed in color-coded rounded rectangles, each assigned a unique color. Inside each rounded rectangle, the four user traits are depicted within white ellipses, while their subcategories with themes are presented below in colored boxes using varying hues.}
    \label{fig:qua_II}
\end{figure*}



In the "Men" group, participants often focused on providing practical and straightforward responses to CAs' emotions. Typical examples included statements like, \textit{"I adjusted my tone based on the agent's response"} and \textit{"I responded logically, addressing the emotional state as needed"}. This suggests that men tend to approach conversations with a focus on analyzing the situation and offering direct, solution-oriented responses rather than being swayed by the agent’s emotional shifts. Conversely, in the "Women" group, participants tended to show more empathy and emotional alignment with the agent. Some reported: \textit{"I tried to match the agent’s emotion and offer comfort"} or \textit{"I adjusted my responses to reflect understanding and support, especially when the agent was sad"}. This reflects a focus on emotional congruence and building a supportive dialogue. This gender distinction aligns with research showing that men often favor pragmatic, problem-solving strategies, while women prioritize empathy and emotional connection with CAs.

\textcolor{black}{\textbf{\textit{Personality}}: \textbf{Personality drives emotional engagement: extraverts mirror actively, ambiverts balance, and introverts engage thoughtfully.}}

Ambiverts demonstrated a flexible and balanced approach to emotional adjustments, adapting naturally to the conversation without being overly reserved or expressive. One ambivert noted, \textit{"I responded based on my unique emotional response"}, showing their ability to engage intuitively without overthinking. Introverts favored a more cautious and reflective style, emphasizing calmness and careful listening. As one introvert shared: \textit{"I tried to be calm and listen carefully so that I could respond appropriately"}, reflecting their tendency to process emotions internally and offer thoughtful, measured responses. Extraverts, in contrast, engaged dynamically with the agent’s emotions, aiming for alignment. One stated: \textit{"I tried to match their emotion as much as I could"}, indicating a focus on maintaining emotional synchronicity and fostering a lively dialogue. These patterns highlight that ambiverts balance engagement, introverts concentrate on reflection, and extraverts prioritize emotional alignment and responsiveness.

\textcolor{black}{\textbf{\textit{Ethnicity}}: \textbf{Ethnicity shapes engagement: Blacks prioritize empathy, Whites show varied involvement, and Mixed adapt based on comfort and cues.}}

Black participants generally emphasized empathy and understanding, adjusting their responses to relate to the agent's emotions. For instance, a typical response was: \textit{"I tried to empathize with the agent and show I understood their feelings"}, highlighting their focus on emotional connection and support. White participants tended to balance logical and emotional adjustments, focusing on maintaining a smooth conversational flow while being mindful of emotional cues. \textit{"I adjusted my response based on the agent's tone, trying to keep the conversation flowing smoothly"}, reflected by one participant, showcasing their blend of practicality and emotional awareness. Mixed often displayed flexibility, adapting to both the tone and emotions of the agent. One stated: \textit{"I changed my response depending on the agent’s mood, trying to match their energy"}, demonstrating their ability to shift strategies depending on the situation.

\textcolor{black}{\textbf{\textit{Geographic Background}}: \textbf{Geography with responses: Europeans use pragmatic/humorous styles, Africans favor empathetic listening and strong engagement.}}

Participants from European countries often adapt their responses by prioritizing emotional balance and maintaining smooth conversational flow. For example, a participant claimed: \textit{"I tried to match the agent’s tone and keep the conversation upbeat"}, reflecting a focus on emotional alignment and politeness. In Asian countries, participants typically focus on respect, harmony, and emotional sensitivity when adjusting their responses. One expressed: \textit{"I tried to respond in a way that would not offend the agent and maintain harmony"}, highlighting a cultural preference for maintaining peace and avoiding conflict. People from African countries emphasized empathy and community-oriented communication. An example was presented: \textit{"I tried to show understanding and offer support in response to the agent’s emotions"}, showing a strong focus on mutual understanding and supportiveness. We found that emotional flow was more revealed in European countries, harmony and respect in Asian countries, empathy and emotional support in African countries.

\subsubsection{Engaged Strategies}
\leavevmode \\ \hspace*{.8em} 
\textcolor{black}{\textbf{\textit{Gender}}: \textbf{Gender for responses: men lean towards structured, practical solutions; women emphasize empathy, personalization, and emotional mirroring.}}

For strategies with female-voiced CAs, men and women also showed nuance. Men leaned towards practical, solution-focused responses, minimizing emotional engagement. For example, when faced with \textit{angry} or \textit{fear}, they offered advice or guidance to resolve the issue. In \textit{happy} or \textit{neutral} situations, they kept their responses straightforward and light, such as saying: \textit{"I responded calmly and gave advice to help the agent move past their \textit{angry}"}, reflecting a controlled, logical approach. Women, on the other hand, engaged more emotionally. When the agent expressed sadness or \textit{fear}, they offered comfort and empathy. In \textit{happy} situations, they mirrored the agent’s joy, focusing on creating a supportive connection. A typical response was: \textit{"I tried to make the agent feel better by showing empathy for their \textit{fear}"}, emphasizing emotional engagement. These differences show that men prioritize logic and resolution, while women emphasize emotional connection and empathy in their responses.

\textcolor{black}{\textbf{\textit{Personality}}: \textbf{Personality shapes responses: extraverts amplify emotions; ambiverts balance empathy; introverts stay calm and thoughtful.}}

Ambiverts maintained a balanced approach, adapting to the agent’s emotions without overcommitting. They switched between engagement and neutrality depending on the situation. For example, \textit{"I responded based on how I felt, keeping things upbeat but not forcing it"}, said by an ambivert, showing flexibility in both positive and negative emotional states. Introverts, by contrast, preferred calm and reflective responses, often taking time to process emotions before engaging. In \textit{neutral} situations, they maintained a reserved tone, with one stating: \textit{"I tried to listen carefully and provide a calm, thoughtful response"}. In emotional situations like \textit{fear} or sadness, introverts offered quiet support, focusing on measured responses. Extraverts displayed a dynamic approach, actively matching the agent’s emotions. In \textit{happy} situations, they mirrored enthusiasm, with one reporting: \textit{"I shared the agent’s excitement and kept the conversation lively"}. Even in negative scenarios, extraverts engaged energetically, using humor or advice to diffuse tension.

\textcolor{black}{\textbf{\textit{Ethnicity}}: \textbf{Ethnicity shapes responses: Blacks are highly empathetic, Whites vary in engagement, and Mixed balance support with practicality.}}

Distinct patterns emerged in responsive strategies based on ethnicity again. Black participants consistently prioritized empathy and emotional support, focusing on understanding the agent’s emotions and providing comfort. In sadness or \textit{fear}, they often offered reassurance. One conveyed: \textit{"I tried to show understanding and offer words of comfort"}, emphasizing emotional care. Even in \textit{neutral} or \textit{happy} interactions, they maintained a warm, supportive tone to sustain a positive atmosphere. White participants adopted a balanced approach, blending emotional engagement with practicality. In scenarios involving \textit{angry} or \textit{fear}, they prioritized de-escalation, offering advice and maintaining calm, as one noted: \textit{"I have tried to address the agent’s concerns while keeping the conversation on track"}. In positive scenarios, they engaged but remained composed. Mixed ethnicity participants demonstrated flexibility, adapting their responses dynamically to the agent’s emotional cues. In positive states, they mirrored enthusiasm, for instance, one mentioned: \textit{"I shared their joy and kept the energy high"}, while in negative states, they balanced empathy with practical advice, aligning with the agent’s mood to offer support".

\textcolor{black}{\textbf{\textit{Geographic Background}}: \textbf{Geographical background shapes responses: Europeans emphasize alignment and practicality, while Africans prioritize empathy and support.}}

Cultural background provided a regional perspective. In European countries, participants balanced emotional engagement with practical problem-solving. They used straightforward responses in \textit{neutral} situations, mirrored joy during \textit{happy} moments, and offered calm, logical advice in \textit{angry} or \textit{fear}, showcasing a pragmatic approach that blends emotional alignment with solution-focused dialogue. In Asia, the emphasis was on harmony and emotional restraint. Participants were polite and reserved in \textit{neutral} or \textit{happy} interactions, and in \textit{angry} or \textit{fear}, they opted for calm, neutral responses, reflecting a preference for maintaining peace and avoiding escalation. In African countries, participants prioritized empathy and emotional connection, offering support and encouragement in \textit{sad} or \textit{fear} scenarios and sharing in joy during \textit{happy} moments, highlighting a communal and deeply engaged approach. These differences underscore how cultural norms shape responsive strategies.


\subsubsection{Potential Strategic Modification}
\leavevmode \\ \hspace*{.8em} 
\textcolor{black}{\textbf{\textit{Gender}}: \textbf{Gender affects strategy adjustments: women are more open to enhancing empathy, while men prefer to keep initial approaches with minor tweaks.}}

Men generally tended to stick to their original strategies, focusing on consistency and practicality. A common response from men was: \textit{"No, I wouldn’t change my approach, I feel it worked well"}, reflecting confidence in their problem-solving abilities and belief that their initial responses were effective. On the other hand, women are more likely to modify their strategies, often seeking to enhance empathy and emotional alignment. One example was: \textit{"Yes, I would modify my strategy to offer more emotional support"}, indicating a greater willingness to adjust based on the dynamics of the conversation. This implies that men’s preference is maintaining practical solutions, while women prioritize emotional adaptability and connection.

\textcolor{black}{\textbf{\textit{Personality}}: \textbf{Personality influences change: extraverts seek more empathy, ambiverts balance contentment with tweaks, and introverts prefer subtle adjustments.}}

Regarding female-voiced CAs, Ambiverts, known for their adaptability, were inclined towards minor adjustments rather than major changes, often expanding on their responses based on the agent’s emotions. An ambivert stated: \textit{"I wouldn’t change much, but I might expand on my responses"}, highlighting their balanced approach. Introverts, in contrast, were more reserved about modifications, believing their thoughtful, measured communication style was effective. One introvert claimed: \textit{"No, I think my approach was appropriate"}, emphasizing their preference for consistency. Extraverts, however, were the most willing to adjust, aiming to enhance engagement and emotional responsiveness. An extravert remarked: \textit{"Yes, I would modify my strategy to be more engaging"}, showcasing their enthusiasm for dynamic interactions. In summary, ambiverts favor small tweaks, introverts prioritize consistency, and extraverts are keen on changes that boost emotional engagement.

\textcolor{black}{\textbf{\textit{Ethnicity}}: \textbf{Ethnicity shapes strategy: Blacks aim to boost empathy, while Whites are generally satisfied but open to enhancing trust and engagement.}}

In terms of strategic modifications, Black participants were generally open to modifying their strategies to deepen emotional connection and empathy, often emphasizing support and community values. A typical response was: \textit{"Yes, I would modify my strategy to be more empathetic and offer more encouragement"}. White participants took a more pragmatic approach, focusing on modifications that would enhance conversation flow or clarity rather than emotional adjustments. One noted: \textit{"I don’t think I would change much, but I might adjust to improve clarity"}. Mixed-ethnicity participants balanced emotional engagement with practicality, showing flexibility in their responses. A common statement was: \textit{"I would adjust my strategy to better align with the agent’s emotions and maintain the conversation’s flow"}, reflecting a dynamic blend of empathy and effective communication.

\textcolor{black}{\textbf{\textit{Geographic Background}}: \textbf{Geographical background matters: Africans prioritize empathy; Europeans prefer current approaches with slight emotional adjustments.}}

European participants preferred a pragmatic approach, enhancing conversational flow and problem-solving, as reflected by a participant who noted: \textit{"I wouldn’t change much, but I might adjust to make the conversation smoother"}, underscoring a focus on clarity and efficiency. The few Asian participants emphasized harmony and emotional restraint, as we received one answer: \textit{"I would modify my strategy to maintain a calm tone and avoid causing offense"}. In contrast, Africans were more open to modifying strategies to foster emotional connection, as seen in a Nigerian participant's statement: \textit{"Yes, I would modify my strategy to be more encouraging and empathetic"}, reflecting a community-oriented and supportive approach.

%% file: Sections/Key_Findings.tex
\begin{table}[!t]
\centering
\small
\caption{Qualified user interaction \textbf{\textit{strategic differences with male- vs. female-voiced CAs}} across user traits.}
\label{tab:difference_I_II}
\begin{tabular}{|p{4.2cm}|p{4.8cm}|p{5.8cm}|}
\hline
\rowcolor{headercolor} \textcolor{white}{\textbf{\textcolor{white}{Qualitative} Metric}} & \textcolor{white}{\textbf{Male-Voiced CAs}} & \textcolor{white}{\textbf{Female-Voiced CAs}} \\
\hline
\rowcolor{darkblue} \textcolor{white}{\textbf{Adjustment of Responses}} & &  \\
\hline
\rowcolor{cyan!20} \textbf{\textit{Gender}} &  & \\
\hline
\rowcolor{white}\textbf{Men} & Prefer practical, structured responses. & Emphasis on tone and flow. \\
\rowcolor{lightgray} \textbf{Women} & Empathy-driven interactions. & Emotional mirroring and personal connection. \\
\hline
\rowcolor{cyan!20} \textbf{\textit{Personality}} &  &  \\
\hline
\rowcolor{white} \textbf{Extraverts} & Active and responsive. & Deeper empathetic engagement. \\
\rowcolor{lightgray} \textbf{Ambiverts} & Contextually balanced responses. & Flexible adaptation to situational changes. \\
\rowcolor{white} \textbf{Introverts} & Prefer reflective responses. & Maintain cautious engagement. \\
\hline
\rowcolor{cyan!20} \textbf{\textit{Ethnicity}} &  &  \\
\hline
\rowcolor{white} \textbf{Black} & Seek empathy and tone alignment. & Expect emotional support. \\
\rowcolor{lightgray} \textbf{White} & View CAs Neutral with skepticism. & View CAs mixed empathy and humor. \\
\rowcolor{white} \textbf{Mixed} & Find CAs adaptable. & Experience balanced engagement. \\
\hline
\rowcolor{cyan!20} \textbf{\textit{Geographic Background}} &  &  \\
\hline
\rowcolor{white} \textbf{European} & Look for flow. & Appreciate varied approaches deom CAs. \\
\rowcolor{lightgray} \textbf{African} & Seek reassurance. & Empathetic tone adjustments. \\
\hline
\rowcolor{darkblue} \textcolor{white}{\textbf{Engaged Strategies}} &  &  \\
\hline
\rowcolor{cyan!20} \textbf{\textit{Gender}} &  & \\
\hline
\rowcolor{white} \textbf{Men} & Practical, neutral engagement. & Enjoy structured humor. \\
\rowcolor{lightgray} \textbf{Women} & Seek empathy. & Find deep emotional mirroring detached. \\
\hline
\rowcolor{cyan!20} \textbf{\textit{Personality}} &  &  \\
\hline
\rowcolor{white} \textbf{Extraverts} & Respond with High energy. & Dynamic with strong empathy. \\
\rowcolor{lightgray} \textbf{Ambiverts} & Balance of empathy and humor. & Adapt to flow. \\
\rowcolor{white} \textbf{Introverts} & Prefer calm, neutral CAs. & Favor thoughtful responses; feel less intense. \\
\hline
\rowcolor{cyan!20} \textbf{\textit{Ethnicity}} & &   \\
\hline
\rowcolor{white} \textbf{Black} & View male CAs as neutral with humor. & Find female CAs emotionally detached. \\
\rowcolor{lightgray} \textbf{White} & Seek deep engagement. & Expect consistent support . \\
\rowcolor{white} \textbf{Mixed} & Experience minimal adjustments. & Find routine responses. \\
\hline
\rowcolor{cyan!20} \textbf{\textit{Geographic Background}} &  & \\
\hline
\rowcolor{white} \textbf{European} & Look for balance. & appreciate natural adaptations. \\
\rowcolor{lightgray} \textbf{African} & Seek calm empathy. & Value reflective support. \\
\hline
\rowcolor{darkblue} \textcolor{white}{\textbf{Potential Strategic Modifications}} &  &  \\
\hline
\rowcolor{cyan!20} \textbf{\textit{Gender}} & &  \\
\hline
\rowcolor{white} \textbf{Men} & Prefer minor tweaks. & Seek enhanced empathy. \\
\rowcolor{lightgray} \textbf{Women} & Look for clarity improvements. & Desire deeper alignment. \\
\hline
\rowcolor{cyan!20} \textbf{\textit{Personality}} &  & \\
\hline
\rowcolor{white} \textbf{Extraverts} & Aim for better engagement. & Seek deeper empathy. \\
\rowcolor{lightgray} \textbf{Ambiverts} & Notice minor tweaks. & Prefer expanded strategies. \\
\rowcolor{white} \textbf{Introverts} & Maintain current approaches. & Seek slight improvements. \\
\hline
\rowcolor{cyan!20} \textbf{\textit{Ethnicity}} & &  \\
\hline
\rowcolor{white} \textbf{Black} & Desire empathetic changes. & Seek greater support. \\
\rowcolor{lightgray} \textbf{White} & Look for balance. & Desire improved trust. \\
\rowcolor{white} \textbf{Mixed} & Experience balanced adjustments. & Seek adaptable changes. \\
\hline
\rowcolor{cyan!20} \textbf{\textit{Geographic Background}} &  &  \\
\hline
\rowcolor{white} \textbf{European} & Expect minor improvements. & Appreciate responsive adaptations. \\
\rowcolor{lightgray} \textbf{African} & Seek better emotional alignment. & Value reassurance and empathy. \\
\hline
\end{tabular}
\end{table}

\section{Key Findings}
\label{find}

\subsection{Strategic Differences Between Male- vs. Female-Voiced CAs}
We dissected distinct user interaction strategic patterns influenced by user traits -- gender, personality, ethnicity, and geographic background -- regarding the qualitative metrics of male-/female-voiced CAs, as shown in Tab. ~\ref{tab:difference_I_II}.

\textbf{Adjustment of Responses:} Males interacting with male CAs favored practical, structured responses, while their interactions with female CAs focused more on adjusting tone and flow. Females exhibited empathy-driven adjustments with male CAs, focusing on emotional mirroring and personal connection with female CAs. Extraverts were active and responsive with male CAs but sought deeper empathetic engagement with female CAs. Ambiverts displayed balanced responses across both CA types, adapting contextually to male CAs and showing flexible situational adjustments with female CAs. Introverts favored reflective and cautious engagement, maintaining calm and neutral responses with both CA types. Ethnicity played a notable role, with Black respondents seeking empathy and tone alignment with male CAs and expecting emotional support from female CAs. White respondents approached male CAs with skepticism, whereas their interactions with female CAs showed a combination of empathy and pragmatic engagement.

\textbf{Engaged Strategies:} Engagement strategies differed notably as well. Males generally employed practical and neutral approaches with male CAs while favoring structured humor and a more detached style of interaction with female CAs. Women consistently sought empathy and deeper emotional mirroring, regardless of CA gender. Extraverts responded with high energy to male CAs and adopted a dynamic, strongly empathetic approach with female CAs. Ambiverts balanced their engagement with humor and empathy for male CAs, showing adaptability in flow when interacting with female CAs. Introverts preferred calm and neutral interactions with male CAs and favored thoughtful, less intense responses with female CAs. Ethnically, Black respondents approached male CAs with neutrality and humor, expecting consistent support from female CAs. In contrast, White respondents demonstrated deeper engagement and a blend of emotional alignment with both CA types.

\textbf{Potential Strategic Modifications:} Modifications were generally minor but reflected a desire for enhanced empathy and emotional alignment. Men sought minor tweaks and clarity improvements, while women desired deeper emotional alignment, particularly with female CAs. Extraverts aimed for better engagement with male CAs and sought expanded empathetic strategies with female CAs. Ambiverts noticed minor areas for improvement across both CA types, showing a need for situational adjustments. Introverts largely maintained their current approaches, emphasizing caution and reflective support, particularly with female CAs. Ethnic differences highlighted that Black respondents desired more empathetic changes, especially with female CAs, while White respondents focused on maintaining balance and trust improvements.

These findings suggest that user traits significantly affect how individuals interact with male/female CAs, with notable preferences for empathy, tone matching, and practical responses shaped by personality, ethnicity, and cultural background.

\subsection{User Interactions with Different Emotional CAs}
Furthermore, we disentangled the nuances of user interactions across user traits in relation to the five emotional states -- \textit{neutral}, \textit{sad}, \textit{angry}, \textit{happy}, and \textit{fear} -- during interactions with both male and female CAs. However, due to the information limitation in the collected data, we were able to extract insights for only a subset of user traits. The summarized findings are presented in Tab. ~\ref{tab:user_inter}.

\begin{table}[!t]
\centering
\small
\caption{User interactions with the \textbf{\textit{five different emotional CAs}} across user traits.}
\label{tab:user_inter}
\begin{tabular}{|p{1.1cm}|p{4.3cm}|p{4.4cm}|p{4.4cm}|}
\hline
\rowcolor{headercolor} \textcolor{white}{\textbf{Emotion}} & \textcolor{white}{\textbf{Male-Voiced CAs}} & \textcolor{white}{\textbf{Female-Voiced CAs}} & \textcolor{white}{\textbf{Both CAs}} \\
\hline
\rowcolor{cyan!20} \textbf{Neutral} & 
\textbf{Men:} Task-focused, practical strategies. \newline
\textbf{Introverts:} Calm, logical engagement. \newline
\textbf{European:} Balanced, structured responses. &
\textbf{Women:} Focused on maintaining engagement. \newline
\textbf{African:} Deeper emotional connection. \newline
\textbf{Introverts:} Occasionally disengaged. &
Generally prompted structured, emotion-matching responses, but often \textbf{led to disengagement} due to lack of emotional connection. \\
\hline
\rowcolor{white} \textbf{Sad} & 
\textbf{Extraverts:} Active emotional support. \newline
\textbf{Men:} Calm reassurance. \newline
\textbf{Black:} High empathy and emotional connection. &
\textbf{Women:} Personal and empathetic engagement. \newline
\textbf{African:} Strong reassurance-driven strategies. \newline
\textbf{Extraverts:} High emotional alignment. &
Elicited \textbf{the highest empathy and emotional alignment} across users, especially with extraverts and ambiverts. \\
\hline
\rowcolor{cyan!20} \textbf{Angry} & 
\textbf{Men:} Calm and effective de-escalation. \newline
\textbf{Extraverts:} Energetic but controlled engagement. \newline
\textbf{White:} Skeptical, detached responses. &
\textbf{Women:} Focus on calming tone but struggle with alignment. \newline
\textbf{Introverts:} Reserved and hesitant. \newline
\textbf{Mixed Ethnicity:} Balanced but emotionally cautious. &
\textbf{Least preferred emotion}; required de-escalation and calm strategies, which male CAs handled more effectively than female CAs. \\
\hline
\rowcolor{white} \textbf{Happy} & 
\textbf{Ambiverts:} Balance humor and engagement. \newline
\textbf{Mixed Ethnicity:} Neutral and steady responses. \newline
\textbf{Extraverts:} Responsive and dynamic. &
\textbf{Women:} Strong emotional mirroring. \newline
\textbf{Extraverts:} High energy and alignment. \newline
\textbf{African:} Deeply empathetic and expressive. &
\textbf{The most preferred emotion}; facilitated high engagement and emotional alignment, particularly with extraverts and ambiverts. \\
\hline
\rowcolor{cyan!20} \textbf{Fear} & 
\textbf{Introverts:} Reserved, reflective strategies. \newline
\textbf{Men:} Structured and logical reassurance. \newline
\textbf{African:} Strong reflection and emotional support. &
\textbf{Women:} Emotionally intense empathy. \newline
\textbf{Extraverts:} High engagement but cautious alignment. \newline
\textbf{Ambiverts:} Adaptable to changing cues. &
Balanced responses combining calm reassurance and empathy, with \textbf{introverts and Africans showing the strongest reflective support}. \\
\hline
\end{tabular}
\end{table}

With \textbf{male CAs}, users often leaned towards practical, solution-focused strategies, especially when emotions like neutrality, anger, and fear were expressed. Men tended to match the male CA’s neutrality and used structured approaches, while women focused more on maintaining engagement, though they sometimes felt disconnected. Extraverts thrived in keeping the conversation lively, regardless of the CA’s emotional state, while introverts preferred calmer, more reflective interactions, particularly with male CAs. Ethnic and geographic differences also influenced response strategies, with Black and African participants generally showing deeper empathy and emotional connection, even when interacting with a \textit{neutral} or \textit{sad} CA.

In contrast, \textbf{female CAs} elicited stronger empathetic and personalized responses, especially in emotions such as \textit{sad} and \textit{happy}. Women showed a greater tendency to emotionally connect with female CAs, reflecting more personalization and empathy, while men continued to apply practical adjustments but struggled more with maintaining engagement. Extraverts responded actively and matched the emotional energy of female CAs, particularly in \textit{happy} and \textit{sad} scenarios, highlighting a preference for these emotionally rich interactions. Ambiverts displayed notable adaptability, balancing empathy and practicality, making them well-suited to navigating the dynamic cues of female CAs. Introverts, however, remained more reserved and sometimes found female CAs’ emotional expressions overwhelming, favoring the steadiness of male CAs.

Across \textbf{both CAs}, \textit{happy} and \textit{sad} were the emotions where users showed the most engagement and emotional alignment. These emotions allowed participants to express empathy and support naturally, aligning well with the interactive strengths of both extraverts and ambiverts. Conversely, \textit{neutral} and \textit{angry} were less preferred, often leading to disengagement or difficulty in connecting, particularly with female CAs. Neutrality frequently prompted structured, emotion-matching responses, while \textit{angry} necessitated de-escalation and calm approaches, which some users, especially introverts and men, managed more effectively with male CAs. By combining the quantitative results from both studies, we found that \textit{happy} emerged as the most preferred emotion, both in terms of emotional engagement and agent awareness. In contrast, the \textit{angry} emotion unanimously ranked as the least desirable one across all dimensions.

Overall, the interaction dynamics reveal that users are generally more engaged with expressive emotions like \textit{happy} and sad, particularly with female CAs, which fostered deeper connections. Male CAs were better suited to interactions involving calm, structured, or rational responses, aligning well with emotions like neutrality and fear. These insights highlight how voice and emotional context play crucial roles in shaping user strategies and engagement levels with CAs.

%% file: Sections/Discussion.tex
\section{Discussion}
\label{dis}

Our study comprehensively explores how users engage with emotion-aware CAs across various emotional scenarios. 
\textcolor{black}{Existing research has largely focused on empathic emotion-responsive strategies~\cite{chin2020empathy,yalccin2020empathy} or generalized user behaviour~\cite{ehtesham2024emobot,lee2023understanding}, often overlooking the dynamic and adaptive strategies users employ when engaging with emotionally expressive agents. Our work bridges this gap by presenting a multi-faceted exploration of user-CA interactions, emphasizing personalization and emotional adaptability.}
By answering the \textbf{RQs} 1 and 2 and connecting user traits with interaction strategies, we understand how emotion-aware CAs can be designed to facilitate more effective, personalized, and emotionally intelligent interactions. Our findings emphasize the importance of emotional alignment between users and CAs, showing how emotion-aware CAs significantly influences users' interaction strategies and experience. Our study sets a foundation for future directions in CA design and development.

\subsection{Positive Emotions Play a Critical Role in Enhancing User Engagement and Satisfaction}
One key finding from our study is the pivotal role emotional context plays in shaping user engagement with emotional-aware CAs. Across both studies, interactions with emotional-aware CAs displaying positive emotions, particularly happiness, were rated as the most comfortable, consistent, and engaging. Users reported a stronger sense of connection and a higher willingness to continue conversations with positive agents, highlighting the importance of positive affect in driving user satisfaction~\cite{mcduff2019longitudinal}. In contrast, negative emotions like sadness and anger led to lower comfort and emotional alignment, making users less inclined to engage deeply in those scenarios.

This finding may significantly inspire the design of emotion-aware CAs. Although emotional intelligence is crucial, not all emotions equally benefit user satisfaction. Emotional-aware CAs that mirror or amplify negative emotions risk disengaging users, especially if the responses feel excessive or misaligned. Thus, responses to CAs for negative emotions should be carefully implemented to maintain comfort and avoid overwhelming users~\cite{diederich2019emulating}. Emotional mirroring is most effective when it aligns with users' emotional states and goals, enhancing engagement and satisfaction~\cite{reeves1996media}. However, overuse of emotional feedback, especially in negative contexts, can reduce the perceived human-like quality of the interaction and diminish user trust~\cite{picard2003affective}. Thus, emotional responses must be contextually aware and thoughtfully managed for an optimal user experience.

\subsection{Gender as the Dominant Factor in Emotional Interaction with CAs}
Our study shows that user traits strongly influence emotional interactions with emotional-aware CAs, with gender differences being the most pronounced. Female users engaged more deeply with the emotional content during interaction. They exhibited a stronger preference for emotionally supportive and empathetic interaction strategies. They tended to offer more comfort and empathy to the CAs, particularly in negative emotional contexts such as sadness and fear. This aligns with research on gendered communication styles, showing that women are more likely to adopt relational and emotionally expressive strategies in human-human and human-agent interactions~\cite{tannen1990gender}. 

On the contrary, male participants tended to prioritize solution-oriented communication, focusing on emotional detachment and practical problem-solving. In negative emotional contexts, their interactions were aimed more at de-escalating and resolving issues rather than engaging deeply with emotional content. This aligns with previous studies showing gender differences in processing emotional information and communication preferences in emotionally charged situations~\cite{mulac2001empirical}. These findings suggest that adapting emotional-aware CAs design to account for gender-specific preferences, such as offering practical or emotionally supportive conversational modes, could enhance user satisfaction and emotional engagement.

\subsection{Personality as a Modulator of Interaction Strategies}

Personality also significantly influenced interaction strategies. Extroverts were highly emotionally engaged, often amplifying or mirroring the CA’s emotional state, especially in positive scenarios like happiness. They preferred dynamic, lively interactions, consistent with their general communication style~\cite{mccrae2004contemplated}. In contrast, introverts favoured calm, reflective interactions, particularly in emotionally intense situations, showing a preference for maintaining emotional distance. Ambiverts took a more flexible approach, blending emotional engagement with practical considerations and adapting their style based on the conversation. These findings highlight the importance of considering personality in CA design, as personality-based customization can make interactions more natural and personalized for diverse users.

\subsection{Cultural Sensitivity in Emotional Engagement}
Cultural background also played a key role in shaping emotional interaction strategies. Participants European countries, particularly balance emotional engagement with pragmatic communication, reflecting cultural norms emphasizing directness and clarity~\cite{hall1976classification}. These users responded to emotionally charged situations by combining emotional alignment with practical problem-solving, especially when interacting with CAs displaying positive emotions. This suggests that CAs for European markets should balance emotional engagement and task-oriented communication, particularly in professional contexts.
In contrast, participants from African countries prioritized empathy and emotional connection in all emotional scenarios. They focused on emotional support and community-oriented communication, especially in response to negative emotions, reflecting cultural norms where relationships and emotional connections are highly valued~\cite{king1991psychological}. CAs for these users should emphasize warmth, empathy, and support, even in neutral or positive interactions. These differences underscore the need for CAs to be culturally sensitive, adapting to diverse communication styles and emotional expressions.

\subsection{Design Inspirations for Emotionally Intelligent CAs}
The findings from this study provide critical guidance for designing the next generation of emotionally intelligent CAs. First and foremost, \textbf{emotional intelligence should be core to CA behavior, and it must be both adaptive and context-aware}. CAs should detect and interpret users’ emotional states in real-time, while dynamically modulating their tone, pitch, and conversational style to align with the context—be it task-focused or emotionally supportive. For instance, structured and neutral tones may be preferred in professional or cognitive tasks, especially by introverts or male users, while emotionally expressive and empathetic voices are more effective in social or affective settings, particularly for extraverts and female users.

Second, \textbf{our findings highlight the importance of personalization in emotional interactions}. User traits—such as gender, personality, ethnicity, and cultural background—significantly shape how people engage with male and female CAs, and with different emotional tones. This calls for CAs capable of tailoring their emotional strategies through real-time user modeling or prior user input. Allowing users to choose between different CA voices (e.g., male or female) or customize the emotional intensity of interactions can further increase user satisfaction and trust. Ambiverts and emotionally adaptable users especially benefit from flexible systems that mirror and adjust to varied emotional cues.

Finally, \textbf{emotionally intelligent CAs must evolve beyond basic emotional recognition}. Current models often detect surface-level emotions like happiness or anger, but real-world interactions involve complex, layered emotions—such as masked frustration, mixed anticipation, or restrained sadness. Systems that can identify and respond to these nuanced affective states will foster more human-like, authentic engagement. This represents a promising direction for future advancements in emotional AI, where the goal is not just to recognize emotions but to relate meaningfully through them.

\subsection{Limitations and Future Research Directions}
While this study offers valuable insights into user interactions with CAs, it has several limitations. The relatively small sample size, despite its diversity, may affect the generalisability of the results. For example, in some metrics, we were limited to analyzing only ‘Black’ and ‘White’ for ethnicity and ‘European’ and ‘African’ for geographic background. Future research should aim to replicate these findings with larger and more varied samples, particularly aiming to include a more diverse participant pool encompassing a wider range of ethnicities, geographic locations, and cultural backgrounds, or alternatively, a larger sample size with balanced distribution across user traits, as these were neither balanced nor controlled in this study. Another notable point is that, despite using Prolific -- an established platform for crowdsourced studies -- and instructing participants to conduct the study in distraction-free environments, it is challenging to fully control environmental factors. Some participants may have completed the study in distracting settings, which could potentially affect the results, given the reliance on hearing the CAs and engaging in dialogues in the study. This issue could be alleviated in future research by involving a significantly larger participant pool or conducting the study with on-site participants in controlled environments. Additionally, our study focused on five emotional scenarios, but CAs in real-world applications will encounter a broader range of emotional states. Future studies should explore how CAs can manage more complex emotions, such as mixed feelings, emotional ambivalence, and prolonged emotional states. Additionally, since our research was divided into two studies based on male and female CAs with different participants, we did not explore how identical individuals might react to different voiced CAs in ways that could influence their subsequent engagement. To address this limitation, future studies could adopt a within-subject design, allowing participants to interact with both male and female CAs in a randomized order, or a counterbalanced design to minimize potential biases. Moreover, while both male and female voices in the CAs were carefully generated for linguistic and emotional accuracy, the asymmetry in their production—using a voice generator for the male voice and a human recording for the female—may introduce potential system bias. Furthermore, while this study focused on single-turn dialogues for consistency and ease of implementation so as to obtain early explorations, our approach did not capture the complexities of multi-turn interactions. Future research could build on this by incorporating more dynamic and progressive multi-turn dialogues to gain deeper insights into the evolution of user strategies over time. Another key area for future research is the integration of multimodal emotional cues. This study primarily examined vocal emotional cues, but real-world emotional communication often involves verbal, visual, and contextual signals. CAs could be improved by incorporating facial recognition, gesture analysis, and contextual understanding to create a more holistic emotional experience. This multimodal approach could enhance the accuracy of emotional recognition and response, making interactions feel more natural and human-like. \textcolor{black}{Lastly,} the rapid advancements in \textcolor{black}{LLMs} and speech emotion synthesis open exciting possibilities for the development of emotion-aware CAs. By incorporating human emotional response strategies, future CAs could engage in more fluid and natural conversations, dynamically adapting to users’ emotional states. This evolution towards more emotionally intelligent CAs could revolutionize the way humans interact with technology, making communication feel not only more personalized but also more empathetic and emotionally attuned.

%% file: Sections/Conclusion.tex
\section{Conclusion}
\label{conc}

Our research probed the evolving role of emotional-aware CAs, focusing on the early explorations of how users engage with them in emotionally charged situations. By analyzing user preferences and strategies, particularly about traits like gender, personality, and cultural background, we identified significant differences in how individuals adapt their interaction strategies across various emotional contexts. The findings emphasize the importance of designing CAs that can handle complex emotional dynamics, providing meaningful and supportive assistance while maintaining user trust and comfort. As AI continues to advance, this research highlights the need for tailored CAs that meet diverse user expectations, contributing to the development of emotionally intelligent CAs that offer inclusive and respectful support for a wide range of users.


%% file: sample-base.bib
@article{ma2022should,
  title={How Should Voice Assistants Deal With Users' Emotions?},
  author={Ma, Yong and Drewes, Heiko and Butz, Andreas},
  journal={arXiv preprint arXiv:2204.02212},
  year={2022}
}

@inproceedings{tolmeijer2021female,
  title={Female by default?--exploring the effect of voice assistant gender and pitch on trait and trust attribution},
  author={Tolmeijer, Suzanne and Zierau, Naim and Janson, Andreas and Wahdatehagh, Jalil Sebastian and Leimeister, Jan Marco Marco and Bernstein, Abraham},
  booktitle={Extended abstracts of the 2021 CHI conference on human factors in computing systems},
  pages={1--7},
  year={2021}
}

@inproceedings{august2024know,
  title={Know Your Audience: The benefits and pitfalls of generating plain language summaries beyond the" general" audience},
  author={August, Tal and Lo, Kyle and Smith, Noah A and Reinecke, Katharina},
  booktitle={Proceedings of the CHI Conference on Human Factors in Computing Systems},
  pages={1--26},
  year={2024}
}

@article{marcus2017measuring,
  title={Measuring emotional response: Comparing alternative approaches to measurement},
  author={Marcus, George E and Neuman, W Russell and MacKuen, Michael B},
  journal={Political Science Research and Methods},
  volume={5},
  number={4},
  pages={733--754},
  year={2017},
  publisher={Cambridge University Press}
}

@article{braun2006using,
  title={Using thematic analysis in psychology},
  author={Braun, Virginia and Clarke, Victoria},
  journal={Qualitative research in psychology},
  volume={3},
  number={2},
  pages={77--101},
  year={2006},
  publisher={Taylor \& Francis}
}

@article{zhang2023redefining,
  title={Redefining qualitative analysis in the AI era: Utilizing ChatGPT for efficient thematic analysis},
  author={Zhang, He and Wu, Chuhao and Xie, Jingyi and Lyu, Yao and Cai, Jie and Carroll, John M},
  journal={arXiv preprint arXiv:2309.10771},
  year={2023}
}

@article{de2024performing,
  title={Performing an inductive thematic analysis of semi-structured interviews with a large language model: An exploration and provocation on the limits of the approach},
  author={De Paoli, Stefano},
  journal={Social Science Computer Review},
  volume={42},
  number={4},
  pages={997--1019},
  year={2024},
  publisher={SAGE Publications Sage CA: Los Angeles, CA}
}

@article{mcdonnell2019chatbots,
  title={Chatbots and gender stereotyping},
  author={McDonnell, Marian and Baxter, David},
  journal={Interacting with Computers},
  volume={31},
  number={2},
  pages={116--121},
  year={2019},
  publisher={Oxford University Press}
}

@article{hertz2021adaptive,
  title={Adaptive empathy: A model for learning empathic responses based on feedback},
  author={Hertz, Uri and Shamay-Tsoory, Simone},
  year={2021},
  publisher={PsyArXiv}
}

@article{rasool2024emotion,
  title={Emotion-Aware Response Generation Using Affect-Enriched Embeddings with LLMs},
  author={Rasool, Abdur and Shahzad, Muhammad Irfan and Aslam, Hafsa and Chan, Vincent},
  journal={arXiv preprint arXiv:2410.01306},
  year={2024}
}

@article{salovey1990emotional,
  title={Emotional intelligence},
  author={Salovey, Peter and Mayer, John D},
  journal={Imagination, cognition and personality},
  volume={9},
  number={3},
  pages={185--211},
  year={1990},
  publisher={Sage Publications Sage CA: Los Angeles, CA}
}

@article{zou2024pilot,
  title={A pilot study of measuring emotional response and perception of LLM-generated questionnaire and human-generated questionnaires},
  author={Zou, Zhao and Mubin, Omar and Alnajjar, Fady and Ali, Luqman},
  journal={Scientific reports},
  volume={14},
  number={1},
  pages={2781},
  year={2024},
  publisher={Nature Publishing Group UK London}
}

@article{mcduff2019longitudinal,
  title={Longitudinal observational evidence of the impact of emotion regulation strategies on affective expression},
  author={McDuff, Daniel and Jun, Eunice and Rowan, Kael and Czerwinski, Mary},
  journal={IEEE Transactions on Affective Computing},
  volume={12},
  number={3},
  pages={636--647},
  year={2019},
  publisher={IEEE}
}

@inproceedings{diederich2019emulating,
  title={Emulating Empathetic Behavior in Online Service Encounters with Sentiment-Adaptive Responses: Insights from an Experiment with a Conversational Agent.},
  author={Diederich, Stephan and Janssen-M{\"u}ller, Max and Brendel, Alfred Benedikt and Morana, Stefan},
  booktitle={ICIS},
  year={2019}
}

@article{hall1976classification,
  title={Classification and ecology of closed-canopy forest in Ghana},
  author={Hall, JB and Swaine, M\_D},
  journal={The Journal of Ecology},
  pages={913--951},
  year={1976},
  publisher={JSTOR}
}

@article{hoy2018alexa,
  title={Alexa, Siri, Cortana, and more: an introduction to voice assistants},
  author={Hoy, Matthew B},
  journal={Medical reference services quarterly},
  volume={37},
  number={1},
  pages={81--88},
  year={2018},
  publisher={Taylor \& Francis}
}

@misc{picard1997ective,
  title={A ective Computing},
  author={Picard, Rosalind W and others},
  year={1997},
  publisher={Citeseer}
}

@article{banziger2012introducing,
  title={Introducing the Geneva Multimodal expression corpus for experimental research on emotion perception.},
  author={B{\"a}nziger, Tanja and Mortillaro, Marcello and Scherer, Klaus R},
  journal={Emotion},
  volume={12},
  number={5},
  pages={1161},
  year={2012},
  publisher={American Psychological Association}
}

@article{edwards2016engaged,
  title={Engaged or frustrated? Disambiguating engagement and frustration in search},
  author={Edwards, Ashlee},
  year={2016}
}

@article{hofstede2001culture,
  title={Culture’s consequences: Comparing values, behaviors, institutions, and organizations across nations},
  author={Hofstede, Geert},
  journal={Thousand Oaks},
  year={2001}
}

@article{matsumoto2006culture,
  title={Culture and cultural worldviews: Do verbal descriptions about culture reflect anything other than verbal descriptions of culture?},
  author={Matsumoto, David},
  journal={Culture \& Psychology},
  volume={12},
  number={1},
  pages={33--62},
  year={2006},
  publisher={Sage Publications London, Thousand Oaks, CA and New Delhi}
}

@article{markus1991cultural,
  title={Cultural variation in the self-concept},
  author={Markus, HR},
  journal={The Self: Interdisplinary approaches/Springer},
  year={1991}
}

@article{bickmore2005social,
  title={Social dialongue with embodied conversational agents},
  author={Bickmore, Timothy and Cassell, Justine},
  journal={Advances in natural multimodal dialogue systems},
  pages={23--54},
  year={2005},
  publisher={Springer}
}

@article{sierros2010durable,
  title={Durable transparent carbon nanotube films for flexible device components},
  author={Sierros, KA and Hecht, DS and Banerjee, DA and Morris, NJ and Hu, L and Irvin, GC and Lee, RS and Cairns, DR},
  journal={Thin Solid Films},
  volume={518},
  number={23},
  pages={6977--6983},
  year={2010},
  publisher={Elsevier}
}

@article{mccrae1992introduction,
  title={An introduction to the five-factor model and its applications},
  author={McCrae, Robert R and John, Oliver P},
  journal={Journal of personality},
  volume={60},
  number={2},
  pages={175--215},
  year={1992},
  publisher={Wiley Online Library}
}

@article{reeves1996media,
  title={The media equation: How people treat computers, television, and new media like real people},
  author={Reeves, Byron and Nass, Clifford},
  journal={Cambridge, UK},
  volume={10},
  number={10},
  pages={19--36},
  year={1996}
}

@misc{nass2005wired,
  title={Wired for speech: How voice activates and advances the human-computer relationship},
  author={Nass, CI},
  year={2005},
  publisher={MIT press}
}

@article{prendinger2005empathic,
  title={THE EMPATHIC COMPANION: A CHARACTER-BASED INTERFACE THAT ADDRESSES USERS'AFFECTIVE STATES},
  author={Prendinger, Helmut and Ishizuka, Mitsuru},
  journal={Applied artificial intelligence},
  volume={19},
  number={3-4},
  pages={267--285},
  year={2005},
  publisher={Taylor \& Francis}
}

@article{zeng2009eigenvalue,
  title={Eigenvalue-based spectrum sensing algorithms for cognitive radio},
  author={Zeng, Yonghong and Liang, Y-C},
  journal={IEEE transactions on communications},
  volume={57},
  number={6},
  pages={1784--1793},
  year={2009},
  publisher={IEEE}
}

@article{clavel2015sentiment,
  title={Sentiment analysis: from opinion mining to human-agent interaction},
  author={Clavel, Chloe and Callejas, Zoraida},
  journal={IEEE Transactions on affective computing},
  volume={7},
  number={1},
  pages={74--93},
  year={2015},
  publisher={IEEE}
}

@inproceedings{higashinaka2014evaluating,
  title={Evaluating coherence in open domain conversational systems.},
  author={Higashinaka, Ryuichiro and Meguro, Toyomi and Imamura, Kenji and Sugiyama, Hiroaki and Makino, Toshiro and Matsuo, Yoshihiro},
  booktitle={INTERSPEECH},
  pages={130--134},
  year={2014}
}

@inproceedings{porcheron2018voice,
  title={Voice interfaces in everyday life},
  author={Porcheron, Martin and Fischer, Joel E and Reeves, Stuart and Sharples, Sarah},
  booktitle={proceedings of the 2018 CHI conference on human factors in computing systems},
  pages={1--12},
  year={2018}
}

@inproceedings{mctear2017rise,
  title={The rise of the conversational interface: A new kid on the block?},
  author={McTear, Michael F},
  booktitle={Future and Emerging Trends in Language Technology. Machine Learning and Big Data: Second International Workshop, FETLT 2016, Seville, Spain, November 30--December 2, 2016, Revised Selected Papers 2},
  pages={38--49},
  year={2017},
  organization={Springer}
}

@article{assunccao2022overview,
  title={An overview of emotion in artificial intelligence},
  author={Assun{\c{c}}{\~a}o, Gustavo and Patr{\~a}o, Bruno and Castelo-Branco, Miguel and Menezes, Paulo},
  journal={IEEE Transactions on Artificial Intelligence},
  volume={3},
  number={6},
  pages={867--886},
  year={2022},
  publisher={IEEE}
}

@article{gu2019model,
  title={A model for basic emotions using observations of behavior in Drosophila},
  author={Gu, Simeng and Wang, Fushun and Patel, Nitesh P and Bourgeois, James A and Huang, Jason H},
  journal={Frontiers in psychology},
  volume={10},
  pages={781},
  year={2019},
  publisher={Frontiers Media SA}
}

@article{tracy2011four,
  title={Four models of basic emotions: A review of Ekman and Cordaro, Izard, Levenson, and Panksepp and Watt},
  author={Tracy, Jessica L and Randles, Daniel},
  journal={Emotion review},
  volume={3},
  number={4},
  pages={397--405},
  year={2011},
  publisher={Sage Publications Sage UK: London, England}
}

@article{king1991psychological,
  title={Psychological, physical, and interpersonal correlates of emotional expressiveness, conflict, and control},
  author={King, Laura A and Emmons, Robert A},
  journal={European Journal of Personality},
  volume={5},
  number={2},
  pages={131--150},
  year={1991},
  publisher={SAGE Publications Sage UK: London, England}
}

@article{mccrae2004contemplated,
  title={A contemplated revision of the NEO Five-Factor Inventory},
  author={McCrae, Robert R and Costa Jr, Paul T},
  journal={Personality and individual differences},
  volume={36},
  number={3},
  pages={587--596},
  year={2004},
  publisher={Elsevier}
}

@article{mulac2001empirical,
  title={Empirical support for the gender-as-culture hypothesis: An intercultural analysis of male/female language differences},
  author={Mulac, Anthony and Bradac, James J and Gibbons, Pamela},
  journal={Human Communication Research},
  volume={27},
  number={1},
  pages={121--152},
  year={2001},
  publisher={Oxford University Press}
}

@article{tannen1990gender,
  title={Gender differences in topical coherence: Creating involvement in best friends' talk},
  author={Tannen, Deborah},
  journal={Discourse processes},
  volume={13},
  number={1},
  pages={73--90},
  year={1990},
  publisher={Taylor \& Francis}
}

@article{picard2003affective,
  title={Affective computing: challenges},
  author={Picard, Rosalind W},
  journal={International Journal of Human-Computer Studies},
  volume={59},
  number={1-2},
  pages={55--64},
  year={2003},
  publisher={Elsevier}
}

@inproceedings{chin2020empathy,
  title={Empathy is all you need: How a conversational agent should respond to verbal abuse},
  author={Chin, Hyojin and Molefi, Lebogang Wame and Yi, Mun Yong},
  booktitle={Proceedings of the 2020 CHI conference on human factors in computing systems},
  pages={1--13},
  year={2020}
}

@article{yalccin2020empathy,
  title={Empathy framework for embodied conversational agents},
  author={Yal{\c{c}}{\i}n, {\"O}zge Nilay},
  journal={Cognitive Systems Research},
  volume={59},
  pages={123--132},
  year={2020},
  publisher={Elsevier}
}

@article{ehtesham2024emobot,
  title={EmoBot: Artificial emotion generation through an emotional chatbot during general-purpose conversations},
  author={Ehtesham-Ul-Haque, Md and D’Rozario, Jacob and Adnin, Rudaiba and Utshaw, Farhan Tanvir and Tasneem, Fabiha and Shefa, Israt Jahan and Al Islam, ABM Alim},
  journal={Cognitive Systems Research},
  volume={83},
  pages={101168},
  year={2024},
  publisher={Elsevier}
}

@article{desmet2018measuring,
  title={Measuring emotion: Development and application of an instrument to measure emotional responses to products},
  author={Desmet, Pieter},
  journal={Funology 2: From Usability to Enjoyment},
  pages={391--404},
  year={2018},
  publisher={Springer}
}

@article{lee2023understanding,
  title={Understanding the empathetic reactivity of conversational agents: Measure development and validation},
  author={Lee, Bumho and Yong Yi, Mun},
  journal={International Journal of Human--Computer Interaction},
  pages={1--19},
  year={2023},
  publisher={Taylor \& Francis}
}

@article{caprara1993big,
  title={The “Big Five Questionnaire”: A new questionnaire to assess the five factor model},
  author={Caprara, Gian Vittorio and Barbaranelli, Claudio and Borgogni, Laura and Perugini, Marco},
  journal={Personality and individual Differences},
  volume={15},
  number={3},
  pages={281--288},
  year={1993},
  publisher={Elsevier}
}

@article{andriella2021have,
  title={Do i have a personality? endowing care robots with context-dependent personality traits},
  author={Andriella, Antonio and Siqueira, Henrique and Fu, Di and Magg, Sven and Barros, Pablo and Wermter, Stefan and Torras, Carme and Alenya, Guillem},
  journal={International Journal of Social Robotics},
  volume={13},
  pages={2081--2102},
  year={2021},
  publisher={Springer}
}

@inproceedings{ma2023emotion,
  title={Emotion-Aware Voice Assistants: Design, Implementation, and Preliminary Insights},
  author={Ma, Yong and Zhang, Yuchong and Bachinski, Miroslav and Fjeld, Morten},
  booktitle={Proceedings of the Eleventh International Symposium of Chinese CHI},
  pages={527--532},
  year={2023}
}

@inproceedings{ma2024understanding,
  title={Understanding Dementia Speech: Towards an Adaptive Voice Assistant for Enhanced Communication},
  author={Ma, Yong and Nordberg, Oda Elise and Zhang, Yuchong and Rongve, Arvid and Bachinski, Miroslav and Fjeld, Morten},
  booktitle={Companion Proceedings of the 16th ACM SIGCHI Symposium on Engineering Interactive Computing Systems},
  pages={15--21},
  year={2024}
}

@inproceedings{kim2019comparing,
  title={Comparing data from chatbot and web surveys: Effects of platform and conversational style on survey response quality},
  author={Kim, Soomin and Lee, Joonhwan and Gweon, Gahgene},
  booktitle={Proceedings of the 2019 CHI conference on human factors in computing systems},
  pages={1--12},
  year={2019}
}

@article{xiao2020tell,
  title={Tell me about yourself: Using an AI-powered chatbot to conduct conversational surveys with open-ended questions},
  author={Xiao, Ziang and Zhou, Michelle X and Liao, Q Vera and Mark, Gloria and Chi, Changyan and Chen, Wenxi and Yang, Huahai},
  journal={ACM Transactions on Computer-Human Interaction (TOCHI)},
  volume={27},
  number={3},
  pages={1--37},
  year={2020},
  publisher={ACM New York, NY, USA}
}

@inproceedings{kumar2010socially,
  title={Socially capable conversational tutors can be effective in collaborative learning situations},
  author={Kumar, Rohit and Ai, Hua and Beuth, Jack L and Ros{\'e}, Carolyn P},
  booktitle={Intelligent Tutoring Systems: 10th International Conference, ITS 2010, Pittsburgh, PA, USA, June 14-18, 2010, Proceedings, Part I 10},
  pages={156--164},
  year={2010},
  organization={Springer}
}

@article{fitzpatrick2017delivering,
  title={Delivering cognitive behavior therapy to young adults with symptoms of depression and anxiety using a fully automated conversational agent (Woebot): a randomized controlled trial},
  author={Fitzpatrick, Kathleen Kara and Darcy, Alison and Vierhile, Molly},
  journal={JMIR mental health},
  volume={4},
  number={2},
  pages={e7785},
  year={2017},
  publisher={JMIR Publications Inc., Toronto, Canada}
}

@inproceedings{lee2019caring,
  title={Caring for Vincent: a chatbot for self-compassion},
  author={Lee, Minha and Ackermans, Sander and Van As, Nena and Chang, Hanwen and Lucas, Enzo and IJsselsteijn, Wijnand},
  booktitle={Proceedings of the 2019 CHI Conference on Human Factors in Computing Systems},
  pages={1--13},
  year={2019}
}

@inproceedings{winkler2020sara,
  title={Sara, the lecturer: Improving learning in online education with a scaffolding-based conversational agent},
  author={Winkler, Rainer and Hobert, Sebastian and Salovaara, Antti and S{\"o}llner, Matthias and Leimeister, Jan Marco},
  booktitle={Proceedings of the 2020 CHI conference on human factors in computing systems},
  pages={1--14},
  year={2020}
}

@article{morris2018towards,
  title={Towards an artificially empathic conversational agent for mental health applications: system design and user perceptions},
  author={Morris, Robert R and Kouddous, Kareem and Kshirsagar, Rohan and Schueller, Stephen M},
  journal={Journal of medical Internet research},
  volume={20},
  number={6},
  pages={e10148},
  year={2018},
  publisher={JMIR Publications Toronto, Canada}
}

@inproceedings{niculescu2014design,
  title={Design and evaluation of a conversational agent for the touristic domain},
  author={Niculescu, Andreea I and Yeo, Kheng Hui and D'Haro, Luis F and Kim, Seokhwan and Jiang, Ridong and Banchs, Rafael E},
  booktitle={Signal and Information Processing Association Annual Summit and Conference (APSIPA), 2014 Asia-Pacific},
  pages={1--10},
  year={2014},
  organization={IEEE}
}

@inproceedings{van2021chatbots,
  title={Chatbots in the tourism industry: The effects of communication style and brand familiarity on social presence and brand attitude},
  author={Van Hooijdonk, Charlotte},
  booktitle={Adjunct Proceedings of the 29th ACM Conference on User Modeling, Adaptation and Personalization},
  pages={375--381},
  year={2021}
}

@article{clark2019state,
  title={The state of speech in HCI: Trends, themes and challenges},
  author={Clark, Leigh and Doyle, Philip and Garaialde, Diego and Gilmartin, Emer and Schl{\"o}gl, Stephan and Edlund, Jens and Aylett, Matthew and Cabral, Jo{\~a}o and Munteanu, Cosmin and Edwards, Justin and others},
  journal={Interacting with computers},
  volume={31},
  number={4},
  pages={349--371},
  year={2019},
  publisher={Oxford University Press}
}

@inproceedings{yang2017perceived,
  title={Perceived emotional intelligence in virtual agents},
  author={Yang, Yang and Ma, Xiaojuan and Fung, Pascale},
  booktitle={Proceedings of the 2017 CHI Conference Extended Abstracts on Human Factors in Computing Systems},
  pages={2255--2262},
  year={2017}
}

@inproceedings{ma2019exploring,
  title={Exploring perceived emotional intelligence of personality-driven virtual agents in handling user challenges},
  author={Ma, Xiaojuan and Yang, Emily and Fung, Pascale},
  booktitle={The World Wide Web Conference},
  pages={1222--1233},
  year={2019}
}

@article{cowan2015voice,
  title={Voice anthropomorphism, interlocutor modelling and alignment effects on syntactic choices in human- computer dialogue},
  author={Cowan, Benjamin R and Branigan, Holly P and Obreg{\'o}n, Mateo and Bugis, Enas and Beale, Russell},
  journal={International Journal of Human-Computer Studies},
  volume={83},
  pages={27--42},
  year={2015},
  publisher={Elsevier}
}

@inproceedings{dahlback2007similarity,
  title={Similarity is more important than expertise: Accent effects in speech interfaces},
  author={Dahlb{\"a}ck, Nils and Wang, QianYing and Nass, Clifford and Alwin, Jenny},
  booktitle={Proceedings of the SIGCHI conference on Human factors in computing systems},
  pages={1553--1556},
  year={2007}
}

@article{dewaele2000personality,
  title={Personality and speech production: A pilot study of second language learners},
  author={Dewaele, Jean-Marc and Furnham, Adrian},
  journal={Personality and Individual differences},
  volume={28},
  number={2},
  pages={355--365},
  year={2000},
  publisher={Elsevier}
}

@article{foster2010user,
  title={User preferences can drive facial expressions: evaluating an embodied conversational agent in a recommender dialogue system},
  author={Foster, Mary Ellen and Oberlander, Jon},
  journal={User modeling and user-adapted interaction},
  volume={20},
  pages={341--381},
  year={2010},
  publisher={Springer}
}

@inproceedings{braun2019your,
  title={At your service: Designing voice assistant personalities to improve automotive user interfaces},
  author={Braun, Michael and Mainz, Anja and Chadowitz, Ronee and Pfleging, Bastian and Alt, Florian},
  booktitle={Proceedings of the 2019 CHI conference on human factors in computing systems},
  pages={1--11},
  year={2019}
}

@inproceedings{andrade2023voice,
  title={A Voice-Assisted Approach for Vehicular Data Querying from Automotive IoT-Based Databases},
  author={Andrade, Matheus and Wanderley, Esther and Azevedo, Mariana and Medeiros, Tha{\'\i}s and Silva, Marianne and Silva, Ivanovitch and Costa, Daniel G},
  booktitle={2023 Symposium on Internet of Things (SIoT)},
  pages={1--5},
  year={2023},
  organization={IEEE}
}

@inproceedings{lei2018insecurity,
  title={The insecurity of home digital voice assistants-vulnerabilities, attacks and countermeasures},
  author={Lei, Xinyu and Tu, Guan-Hua and Liu, Alex X and Li, Chi-Yu and Xie, Tian},
  booktitle={2018 IEEE conference on communications and network security (CNS)},
  pages={1--9},
  year={2018},
  organization={IEEE}
}

@inproceedings{zargham2022want,
  title={“I Want It That Way”: Exploring Users’ Customization and Personalization Preferences for Home Assistants},
  author={Zargham, Nima and Alexandrovsky, Dmitry and Erich, Jan and Wenig, Nina and Malaka, Rainer},
  booktitle={CHI conference on human factors in computing systems extended abstracts},
  pages={1--8},
  year={2022}
}

@inproceedings{yang2021designing,
  title={Designing conversational agents: A self-determination theory approach},
  author={Yang, Xi and Aurisicchio, Marco},
  booktitle={Proceedings of the 2021 CHI Conference on Human Factors in Computing Systems},
  pages={1--16},
  year={2021}
}

@inproceedings{clark2019makes,
  title={What makes a good conversation? Challenges in designing truly conversational agents},
  author={Clark, Leigh and Pantidi, Nadia and Cooney, Orla and Doyle, Philip and Garaialde, Diego and Edwards, Justin and Spillane, Brendan and Gilmartin, Emer and Murad, Christine and Munteanu, Cosmin and others},
  booktitle={Proceedings of the 2019 CHI conference on human factors in computing systems},
  pages={1--12},
  year={2019}
}

@article{feine2019taxonomy,
  title={A taxonomy of social cues for conversational agents},
  author={Feine, Jasper and Gnewuch, Ulrich and Morana, Stefan and Maedche, Alexander},
  journal={International Journal of human-computer studies},
  volume={132},
  pages={138--161},
  year={2019},
  publisher={Elsevier}
}

@inproceedings{candello2017evaluating,
  title={Evaluating the conversation flow and content quality of a multi-bot conversational system},
  author={Candello, Heloisa and Vasconcelos, Marisa and Pinhanez, Claudio},
  booktitle={Proceedings of the 2017 CHI Conference Extended Abstracts on Human Factors in Computing Systems},
  year={2017}
}

@article{hu2022acoustically,
  title={The acoustically emotion-aware conversational agent with speech emotion recognition and empathetic responses},
  author={Hu, Jiaxiong and Huang, Yun and Hu, Xiaozhu and Xu, Yingqing},
  journal={IEEE Transactions on Affective Computing},
  volume={14},
  number={1},
  pages={17--30},
  year={2022},
  publisher={IEEE}
}

@inproceedings{aneja2021understanding,
  title={Understanding conversational and expressive style in a multimodal embodied conversational agent},
  author={Aneja, Deepali and Hoegen, Rens and McDuff, Daniel and Czerwinski, Mary},
  booktitle={Proceedings of the 2021 CHI conference on human factors in computing systems},
  pages={1--10},
  year={2021}
}

@article{scheirer2002frustrating,
  title={Frustrating the user on purpose: a step toward building an affective computer},
  author={Scheirer, Jocelyn and Fernandez, Raul and Klein, Jonathan and Picard, Rosalind W},
  journal={Interacting with computers},
  volume={14},
  number={2},
  pages={93--118},
  year={2002},
  publisher={OUP}
}

@article{overall2014attachment,
  title={Attachment anxiety and reactions to relationship threat: the benefits and costs of inducing guilt in romantic partners.},
  author={Overall, Nickola C and Girme, Yuthika U and Lemay Jr, Edward P and Hammond, Matthew D},
  journal={Journal of personality and social psychology},
  volume={106},
  number={2},
  pages={235},
  year={2014},
  publisher={American Psychological Association}
}

@article{mcquiggan2007modeling,
  title={Modeling and evaluating empathy in embodied companion agents},
  author={McQuiggan, Scott W and Lester, James C},
  journal={International Journal of Human-Computer Studies},
  volume={65},
  number={4},
  pages={348--360},
  year={2007},
  publisher={Elsevier}
}

@article{diederich2022design,
  title={On the design of and interaction with conversational agents: An organizing and assessing review of human-computer interaction research},
  author={Diederich, Stephan and Brendel, Alfred Benedikt and Morana, Stefan and Kolbe, Lutz},
  journal={Journal of the Association for Information Systems},
  volume={23},
  number={1},
  pages={96--138},
  year={2022}
}

@article{gornemann2022emotional,
  title={Emotional responses to human values in technology: The case of conversational agents},
  author={G{\"o}rnemann, Esther and Spiekermann, Sarah},
  journal={Human--Computer Interaction},
  pages={1--28},
  year={2022},
  publisher={Taylor \& Francis}
}

@inproceedings{lee2019does,
  title={" What does your Agent look like?" A Drawing Study to Understand Users' Perceived Persona of Conversational Agent},
  author={Lee, Sunok and Kim, Sungbae and Lee, Sangsu},
  booktitle={Extended abstracts of the 2019 CHI conference on human factors in computing systems},
  pages={1--6},
  year={2019}
}

@inproceedings{jeong2019exploring,
  title={Exploring effects of conversational fillers on user perception of conversational agents},
  author={Jeong, Yuin and Lee, Juho and Kang, Younah},
  booktitle={Extended abstracts of the 2019 CHI conference on human factors in computing systems},
  pages={1--6},
  year={2019}
}

@inproceedings{luger2016like,
  title={" Like Having a Really Bad PA" The Gulf between User Expectation and Experience of Conversational Agents},
  author={Luger, Ewa and Sellen, Abigail},
  booktitle={Proceedings of the 2016 CHI conference on human factors in computing systems},
  pages={5286--5297},
  year={2016}
}

@inproceedings{volkel2021eliciting,
  title={Eliciting and analysing users’ envisioned dialogues with perfect voice assistants},
  author={V{\"o}lkel, Sarah Theres and Buschek, Daniel and Eiband, Malin and Cowan, Benjamin R and Hussmann, Heinrich},
  booktitle={Proceedings of the 2021 CHI conference on human factors in computing systems},
  pages={1--15},
  year={2021}
}

@article{simms2019does,
  title={Does the number of response options matter? Psychometric perspectives using personality questionnaire data.},
  author={Simms, Leonard J and Zelazny, Kerry and Williams, Trevor F and Bernstein, Lee},
  journal={Psychological assessment},
  volume={31},
  number={4},
  pages={557},
  year={2019},
  publisher={American Psychological Association}
}

@article{he2017enhancing,
  title={On enhancing the cross--cultural comparability of Likert--scale personality and value measures: A comparison of common procedures},
  author={He, Jia and Van de Vijver, Fons JR and Fetvadjiev, Velichko H and de Carmen Dominguez Espinosa, Alejandra and Adams, Byron and Alonso--Arbiol, Itziar and Aydinli--Karakulak, Arzu and Buzea, Carmen and Dimitrova, Radosveta and Fortin, Alvaro and others},
  journal={European Journal of Personality},
  volume={31},
  number={6},
  pages={642--657},
  year={2017},
  publisher={SAGE Publications Sage UK: London, England}
}

@inproceedings{gupta2022trust,
  title={To trust or not to trust: How a conversational interface affects trust in a decision support system},
  author={Gupta, Akshit and Basu, Debadeep and Ghantasala, Ramya and Qiu, Sihang and Gadiraju, Ujwal},
  booktitle={Proceedings of the ACM Web Conference 2022},
  pages={3531--3540},
  year={2022}
}

@article{palan2018prolific,
  title={Prolific. ac—A subject pool for online experiments},
  author={Palan, Stefan and Schitter, Christian},
  journal={Journal of behavioral and experimental finance},
  volume={17},
  pages={22--27},
  year={2018},
  publisher={Elsevier}
}

@article{peer2017beyond,
  title={Beyond the Turk: Alternative platforms for crowdsourcing behavioral research},
  author={Peer, Eyal and Brandimarte, Laura and Samat, Sonam and Acquisti, Alessandro},
  journal={Journal of experimental social psychology},
  volume={70},
  pages={153--163},
  year={2017},
  publisher={Elsevier}
}

@book{geary1998male,
  title={Male, female: The evolution of human sex differences},
  author={Geary, David C},
  year={1998},
  publisher={American Psychological Association Washington, DC}
}

@inproceedings{yang2019understanding,
  title={Understanding affective experiences with conversational agents},
  author={Yang, Xi and Aurisicchio, Marco and Baxter, Weston},
  booktitle={proceedings of the 2019 CHI conference on human factors in computing systems},
  pages={1--12},
  year={2019}
}

@article{todd2019demographic,
  title={Demographic and behavioral profiles of nonbinary and binary transgender youth},
  author={Todd, Kieran and Peitzmeier, Sarah M and Kattari, Shanna K and Miller-Perusse, Michael and Sharma, Akshay and Stephenson, Rob},
  journal={Transgender Health},
  volume={4},
  number={1},
  pages={254--261},
  year={2019},
  publisher={Mary Ann Liebert, Inc., publishers 140 Huguenot Street, 3rd Floor New~…}
}

@inproceedings{taylor2024cruising,
  title={Cruising Queer HCI on the DL: A Literature Review of LGBTQ+ People in HCI},
  author={Taylor, Jordan and Simpson, Ellen and Tran, Anh-Ton and Brubaker, Jed R and Fox, Sarah E and Zhu, Haiyi},
  booktitle={Proceedings of the CHI Conference on Human Factors in Computing Systems},
  pages={1--21},
  year={2024}
}

@inproceedings{chaves2018single,
  title={Single or multiple conversational agents? An interactional coherence comparison},
  author={Chaves, Ana Paula and Gerosa, Marco Aurelio},
  booktitle={Proceedings of the 2018 CHI Conference on Human Factors in Computing Systems},
  pages={1--13},
  year={2018}
}

@inproceedings{sciuto2018hey,
  title={" Hey Alexa, What's Up?" A Mixed-Methods Studies of In-Home Conversational Agent Usage},
  author={Sciuto, Alex and Saini, Arnita and Forlizzi, Jodi and Hong, Jason I},
  booktitle={Proceedings of the 2018 designing interactive systems conference},
  pages={857--868},
  year={2018}
}

@inproceedings{cai2022impacts,
  title={Impacts of personal characteristics on user trust in conversational recommender systems},
  author={Cai, Wanling and Jin, Yucheng and Chen, Li},
  booktitle={Proceedings of the 2022 CHI Conference on Human Factors in Computing Systems},
  pages={1--14},
  year={2022}
}

@article{oh2020differences,
  title={Differences in interactions with a conversational agent},
  author={Oh, Young Hoon and Chung, Kyungjin and Ju, Da Young},
  journal={International journal of environmental research and public health},
  volume={17},
  number={9},
  pages={3189},
  year={2020},
  publisher={MDPI}
}

@inproceedings{bickmore2007practical,
  title={Practical approaches to comforting users with relational agents},
  author={Bickmore, Timothy and Schulman, Daniel},
  booktitle={CHI'07 extended abstracts on Human factors in computing systems},
  pages={2291--2296},
  year={2007}
}

@article{ling2021factors,
  title={Factors influencing users' adoption and use of conversational agents: A systematic review},
  author={Ling, Erin Chao and Tussyadiah, Iis and Tuomi, Aarni and Stienmetz, Jason and Ioannou, Athina},
  journal={Psychology \& marketing},
  volume={38},
  number={7},
  pages={1031--1051},
  year={2021},
  publisher={Wiley Online Library}
}

@inproceedings{langevin2021heuristic,
  title={Heuristic evaluation of conversational agents},
  author={Langevin, Raina and Lordon, Ross J and Avrahami, Thi and Cowan, Benjamin R and Hirsch, Tad and Hsieh, Gary},
  booktitle={Proceedings of the 2021 CHI Conference on Human Factors in Computing Systems},
  pages={1--15},
  year={2021}
}

@article{jabir2023evaluating,
  title={Evaluating conversational agents for mental health: scoping review of outcomes and outcome measurement instruments},
  author={Jabir, Ahmad Ishqi and Martinengo, Laura and Lin, Xiaowen and Torous, John and Subramaniam, Mythily and Tudor Car, Lorainne},
  journal={Journal of Medical Internet Research},
  volume={25},
  pages={e44548},
  year={2023},
  publisher={JMIR Publications Toronto, Canada}
}

@inproceedings{wambsganss2020conversational,
  title={A conversational agent to improve response quality in course evaluations},
  author={Wambsganss, Thiemo and Winkler, Rainer and S{\"o}llner, Matthias and Leimeister, Jan Marco},
  booktitle={Extended Abstracts of the 2020 CHI conference on human factors in computing systems},
  pages={1--9},
  year={2020}
}

@article{christoforakos2021connect,
  title={Connect with me. exploring influencing factors in a human-technology relationship based on regular chatbot use},
  author={Christoforakos, Lara and Feicht, Nina and Hinkofer, Simone and L{\"o}scher, Annalena and Schlegl, Sonja F and Diefenbach, Sarah},
  journal={Frontiers in digital health},
  volume={3},
  pages={689999},
  year={2021},
  publisher={Frontiers Media SA}
}

@inproceedings{divekar2019you,
  title={You talkin’to me? A practical attention-aware embodied agent},
  author={Divekar, Rahul R and Kephart, Jeffrey O and Mou, Xiangyang and Chen, Lisha and Su, Hui},
  booktitle={Human-Computer Interaction--INTERACT 2019: 17th IFIP TC 13 International Conference, Paphos, Cyprus, September 2--6, 2019, Proceedings, Part III 17},
  pages={760--780},
  year={2019},
  organization={Springer}
}

@article{levinson1983pragmatics,
  title={Pragmatics},
  author={Levinson, Stephen C},
  journal={Cambridge UP},
  year={1983}
}

@article{schegloff1973opening,
  title={Opening up closings},
  author={Schegloff, Emanuel A},
  journal={Semiotica},
  year={1973}
}

@article{weizenbaum1966eliza,
  title={ELIZA—a computer program for the study of natural language communication between man and machine},
  author={Weizenbaum, Joseph},
  journal={Communications of the ACM},
  volume={9},
  number={1},
  pages={36--45},
  year={1966},
  publisher={ACM New York, NY, USA}
}

@inproceedings{shang2015neural,
  title={Neural Responding Machine for Short-Text Conversation},
  author={Shang, Lifeng and Lu, Zhengdong and Li, Hang},
  booktitle={Proceedings of the 53rd Annual Meeting of the Association for Computational Linguistics and the 7th International Joint Conference on Natural Language Processing (Volume 1: Long Papers)},
  pages={1577--1586},
  year={2015}
}

@inproceedings{chen2019driven,
  title={Driven answer generation for product-related questions in e-commerce},
  author={Chen, Shiqian and Li, Chenliang and Ji, Feng and Zhou, Wei and Chen, Haiqing},
  booktitle={Proceedings of the Twelfth ACM International Conference on Web Search and Data Mining},
  pages={411--419},
  year={2019}
}

@inproceedings{molino2018cota,
  title={Cota: Improving the speed and accuracy of customer support through ranking and deep networks},
  author={Molino, Piero and Zheng, Huaixiu and Wang, Yi-Chia},
  booktitle={Proceedings of the 24th ACM SIGKDD International Conference on Knowledge Discovery \& Data Mining},
  pages={586--595},
  year={2018}
}

@inproceedings{wan2016match,
  title={Match-SRNN: modeling the recursive matching structure with spatial RNN},
  author={Wan, Shengxian and Lan, Yanyan and Xu, Jun and Guo, Jiafeng and Pang, Liang and Cheng, Xueqi},
  booktitle={Proceedings of the Twenty-Fifth International Joint Conference on Artificial Intelligence},
  pages={2922--2928},
  year={2016}
}

@inproceedings{lowe2015ubuntu,
  title={The Ubuntu Dialogue Corpus: A Large Dataset for Research in Unstructured Multi-Turn Dialogue Systems},
  author={Lowe, Ryan and Pow, Nissan and Serban, Iulian and Pineau, Joelle},
  booktitle={Proceedings of the 16th Annual Meeting of the Special Interest Group on Discourse and Dialogue},
  pages={285},
  year={2015},
  organization={Association for Computational Linguistics}
}

@inproceedings{10.2312:evs.20201049,
booktitle = {EuroVis 2020 - Short Papers},
editor = {Kerren, Andreas and Garth, Christoph and Marai, G. Elisabeta},
title = {{Task-based Colormap Design Supporting Visual Comprehension in Process Tomography}},
author = {Zhang, Yuchong and Fjeld, Morten and Said, Alan and Fratarcangeli, Marco},
year = {2020},
publisher = {The Eurographics Association},
ISBN = {978-3-03868-106-9},
DOI = {10.2312/evs.20201049}
}

@article{zhang2021affective,
  title={Affective colormap design for accurate visual comprehension in industrial tomography},
  author={Zhang, Yuchong and Fjeld, Morten and Fratarcangeli, Marco and Said, Alan and Zhao, Shengdong},
  journal={Sensors},
  volume={21},
  number={14},
  pages={4766},
  year={2021},
  publisher={MDPI}
}

@inproceedings{ma2025advancing,
  title={Advancing User-Voice Interaction: Exploring Emotion-Aware Voice Assistants Through a Role-Swapping Approach},
  author={Ma, Yong and Zhang, Yuchong and Fu, Di and Zubicueta Portales, Stephanie and Kragic, Danica and Fjeld, Morten},
  booktitle={International Conference on Human-Computer Interaction},
  pages={303--320},
  year={2025},
  organization={Springer}
}

@inproceedings{zhang2020automated,
  title={Automated microwave tomography (Mwt) image segmentation: State-of-the-art implementation and evaluation},
  author={Zhang, Yuchong and Ma, Yong and Omrani, Adel and Yadav, Rahul and Fjeld, Morten and Fratarcangeli, Marco},
  booktitle={WSCG 2020: Full Papers Proceedings of the 28th International Conference in Central Europe on Computer Graphics, Visualization and Computer Vision},
  pages={126--136},
  year={2020},
  publisher={V{\'a}clav Skala - UNION Agency},
  address={Plze{\v{n}}, Czech Republic},
  doi={10.24132/CSRN.2020.3001.15}
}

@inproceedings{zhang2023industrial,
  title={Is industrial tomography ready for augmented reality? a need-finding study of how augmented reality can be adopted by industrial tomography experts},
  author={Zhang, Yuchong and Nowak, Adam and Rao, Guruprasad and Romanowski, Andrzej and Fjeld, Morten},
  booktitle={International Conference on Human-Computer Interaction},
  pages={523--535},
  year={2023},
  organization={Springer}
}

@inproceedings{zhang2022initial,
  title={An initial exploration of visual cues in head-mounted display augmented reality for book searching},
  author={Zhang, Yuchong and Nowak, Adam and Romanowski, Andrzej and Fjeld, Morten},
  booktitle={Proceedings of the 21st International Conference on Mobile and Ubiquitous Multimedia},
  pages={273--275},
  year={2022}
}
